\documentclass[12pt]{article}
\pdfoutput=1

\usepackage[bulletsep]{collref}
\usepackage{amssymb,graphicx}
\usepackage[intlimits]{amsmath}
\usepackage{bbm}
\usepackage{xcolor}
\usepackage{hyperref}
\usepackage{comment}
\usepackage[small]{subfigure}

\usepackage{tikz}
\usetikzlibrary{matrix,arrows,shapes,decorations.fractals,decorations.pathmorphing,shadows,patterns,decorations.markings}
\definecolor{yellow}{cmyk}{0,0,1,0}
\definecolor{magenta2}{cmyk}{0.25,0.8,0,0.1}
\definecolor{cyan2}{cmyk}{0.9,0.02,0,0.1}
\definecolor{green2}{cmyk}{1,0,1,0.15}
\definecolor{oran}{cmyk}{0,0.2,0.57,0}

\usepackage{MnSymbol}

\makeatletter \@addtoreset{equation}{section} \makeatother

\makeatletter
\let\old@startsection=\@startsection
\let\oldl@section=\l@section
\renewcommand{\@startsection}[6]{\old@startsection{#1}{#2}{#3}{#4}{#5}{#6\mathversion{bold}}}
\renewcommand{\l@section}[2]{\oldl@section{\mathversion{bold}#1}{#2}}
\makeatother

\makeatletter
\let\old@makecaption=\@makecaption
\def\@makecaption{\small\old@makecaption}
\makeatother

\renewcommand{\leq}{\leqslant}

\newcommand{\PU}{\overrightarrow{U}}
\newcommand{\AU}{\overleftarrow{U}}

\begin{document}


\begin{flushright}\footnotesize
\texttt{NORDITA 2018-091} \\
\texttt{UUITP-48/18}
\end{flushright}

\renewcommand{\thefootnote}{\fnsymbol{footnote}}
\setcounter{footnote}{0}

\begin{center}
{\Large\textbf{\mathversion{bold} Dyson equations for correlators \\ of Wilson loops {\color{red}}}
\par}

\vspace{0.8cm}

\textrm{\!\!Diego Correa$^{1}$, Pablo Pisani$^{1}$, Alan Rios Fukelman$^{2}$,
Konstantin Zarembo$^{3,4,5}$\footnote{Also at ITEP, Moscow, Russia}}
\vspace{4mm}

$^{1}$\textit{Instituto de F\'isica La Plata, CONICET, Universidad Nacional de La Plata C.C. 67, 1900, La Plata, Argentina}
\\
$^{2}$\textit{Institut de Ci\`encies del Cosmos, Universitat de Barcelona, Mart\'i i Franqu\`es 1, 08028 Barcelona, Spain}
\\
$^{3}$\textit{Nordita,  Stockholm University and KTH Royal Institute of Technology,
Roslagstullsbacken 23, SE-106 91 Stockholm, Sweden}
\\
$^{4}$\textit{Department of Physics and Astronomy, Uppsala University\\
SE-751 08 Uppsala, Sweden}\\
$^{5}$\textit{Hamilton Mathematics Institute, Trinity College Dublin, Dublin 2, Ireland}\\
\vspace{0.2cm}
\texttt{correa@fisica.unlp.edu.ar,pisani@fisica.unlp.edu.ar,
ariosfukelman@icc.ub.edu, zarembo@nordita.org}

\vspace{3mm}


\par\vspace{1cm}

\textbf{Abstract} \vspace{3mm}

\begin{minipage}{13cm}
By considering a Gaussian truncation of ${\cal N}=4$ super Yang-Mills, we derive a set of Dyson equations that account for the ladder diagram contribution to connected correlators of circular Wilson loops. We consider different numbers of loops, with different relative orientations. We show that the Dyson equations admit a spectral representation in terms of eigenfunctions of a Schr\"odinger problem, whose classical limit describes the strong coupling limit of the ladder resummation. We also verify that in supersymmetric cases the exact solution to the Dyson equations reproduces known matrix model results.

\end{minipage}

\end{center}

\vspace{0.5cm}



\setcounter{page}{1}
\renewcommand{\thefootnote}{\arabic{footnote}}
\setcounter{footnote}{0}

\section{Introduction}

In the study of Wilson loops expectation values and correlators, the ladder diagrams contribution can be separated from the rest simply by identifying Feynman diagrams with no vertices.
Although ladder diagrams only account for observables partially, there are compelling motivations to focus our attention on this particular type of contribution. When restricting to the case of supersymmetric circular Wilson loops, it is possible to argue that all diagrams with vertices cancel each other, ladder approximation becomes exact, and one can obtain exact, non-perturbative results for a number of Wilson loop observables \cite{Erickson:2000af,Drukker:2000rr,Pestun:2007rz} (see \cite{Zarembo:2016bbk} for a review).

Another case when ladder resummation is rigorously justified arises upon analytic continuation in the scalar coupling of the Wilson loop. Scalar ladder diagrams are then enhanced compared to other contributions and their sum constitutes a first order of a systematic expansion \cite{Correa:2012nk}. Apart from a detailed match to string theory at strong coupling, all-order results obtained in this limit feature intriguing connections to integrability \cite{Gromov:2016rrp,Kim:2017sju,Cavaglia:2018lxi}.

In this article we revisit resummation of ladder diagrams for the correlators of circular loops \cite{Zarembo:2001jp,Correa:2018lyl}, in order to clarify some previous results and generalize the analysis in various ways.
Although ladder diagrams do not give the precise answer in this case, their resummation in the planar limit could capture anyway the essential behavior expected from the dual string theory analysis in the strong coupling limit. For example, the ladder contribution to the connected correlator exhibits a phase transition that can be associated with the string breaking phase transition pointed out by Gross and Ooguri \cite{Gross:1998gk}.

The ladder approximation has been analyzed in many ways and for various configurations of Wilson loops \cite{Erickson:1999qv,Erickson:2000af,Zarembo:2001jp,Klebanov:2006jj,Bykov:2012sc,Henn:2013wfa,Marmiroli:2012ny,Bonini:2016fnc,Correa:2018lyl}, providing insight into their behavior at finite 't~Hooft coupling constant $\lambda$, and yielding all-loop results that can be contrasted with the predictions of the AdS/CFT duality in the strong coupling limit.

We will discuss in detail the connected correlator of two co-axial circular Wilson loops, either for the same or opposite spacetime orientations. To account for the ladder contribution, we derive Dyson equations  by a systematic procedure based on Gaussian average over the fields that participate in the Wilson loops. The resulting Dyson equations can be reduced to a Schr\"odinger problem whose classical limit captures the strong coupling limit of the ladder contribution. For Wilson loops of opposite orientation the ladder contribution to the connected correlator exhibits a phase transition resembling the Gross-Ooguri one. We also find supersymmetric critical relations between spacetime and internal space separations \cite{Correa:2018lyl}, such that the ladder contributions can be exactly found and agree with matrix model results from localization.

Finally, we show how to extend this analysis for correlators of more than two loops, by considering the case of three Wilson loops. The system of integral equations turns out to be more intricate in this case. Nevertheless, we can solve it exactly for the critical case, recovering again known matrix model results.

\section{Dyson equations for two loops correlator}

General correlators of Wilson loops are not expected to be fully described by a ladder approximation, since one would be neglecting interaction diagrams that do contribute to the expectation value. Nevertheless, and as it has been shown \cite{Correa:2018lyl}, for certain configurations correlators can be properly described by this reduced set of diagrams allowing, not only an exact match with the dual string theory calculation, but also a description of a phase transition of the Gross-Ooguri type \cite{Gross:1998gk,Zarembo:1999bu,Olesen:2000ji,Kim:2001td}.
Therefore, we begin by deriving an integral Dyson equation whose solutions account for the resummation of ladder diagrams. Our procedure is fairly general and the derivation applies to any Wilson loop correlator, but we will focus on the circular Wilson loop for concreteness.

A  locally supersymmetric Wilson loop in the $\mathcal{N}=4$ SYM theory \cite{Maldacena:1998im}
depends on the representation of the gauge group, which we take to be the fundamental of $U(N)$, the spacetime trajectory $x^\mu(t)$ and the internal space trajectory $n^I(t)$, where $n^I(t)$ is a unit six-component vector at each $t$:
\begin{equation}
 W(C;n^I)= {\rm tr}\, P\exp
 \oint_Cdt\,\left(i A_\mu\dot{x}^\mu  + \Phi _I n^I |\dot{x}|\right).
 \label{WLdef}
\end{equation}

In this work we focus on co-axial circular Wilson loops with constant separation along the symmetry axis and along $S^5$:
\begin{equation}
\!C_a/\bar{C}_a:\quad\! x^\mu_a = (R_a \cos t,\pm R_a\sin t,h_a,0),\quad \! n^I_a=(\cos\gamma_a,\sin\gamma_a,0,0,0,0),
\end{equation}
where the index $a$  labels different loops in a multi-loop correlator.
The contour $\bar{C}_a$ has opposite orientation to $C_a$.

Such configurations of Wilson loops have been studied in the past.
The correlator of two loops of opposite orientation is known perturbatively up to the two-loop order \cite{Plefka:2001bu,Arutyunov:2001hs}. At strong coupling the corresponding minimal surface was found in \cite{Zarembo:1999bu,Olesen:2000ji,Drukker:2005cu}. The general solution in the latter case, that includes separation on $S^{5}$ in addition to arbitrary geometric parameters, was obtained in \cite{Correa:2018lyl}. For the circles of the same orientation the correlator is known at two loops as well \cite{Arutyunov:2001hs}. The connected minimal surface most likely does not exist for parallel circles, as we discuss later in the text.
Non-co-axial circular loops, in particular those sharing a contact point, were also studied recently, both at weak and at strong coupling \cite{Dorn:2018srz}. In this work we concentrate on the contribution of ladder diagrams to co-axial circular loop correlators.

Restriction to ladder diagrams is equivalent to Gaussian integration over
$\Phi _I$ and $A_\mu $, disregarding all interaction terms in the action.
For BPS configurations  of Wilson loops (for instance, for the expectation value of a single circular loop) the Gaussian approximation is actually exact \cite{Pestun:2007rz}. Truncation to ladders can be also justified when the $S^5$  couplings of the Wilson loops are imaginary and very large. In that case  ladders constitute the first order of a systematic expansion in a small parameter \cite{Correa:2012nk}. While in general restriction to ladders is not a systematic approximation, it might capture qualitative features of the exact answer even when not rigorously justified. We will thus treat $\Phi _I$ and $A_\mu $ as free fields from now on. In addition, we will take into account only planar diagrams systematically neglecting $1/N$ corrections.

\begin{figure}[t]
\begin{center}
\begin{tikzpicture}
\draw[green2,very thick,decoration={snake,amplitude=2pt, segment length=4pt}, decorate]  (1,-1) -- (2,-2);
\draw[green2,very thick,decoration={snake,amplitude=2pt, segment length=4pt}, decorate]  (-1,-1) -- (-2,-2);
\draw[green2,very thick, dashed]  (-1,1) -- (-2,2);
\draw[green2,very thick,decoration={snake,amplitude=2pt, segment length=4pt}, decorate]  (-1.17874, 0.781397) -- (-2.35747, 1.56279);
\draw[green2,very thick, dashed]  (1,1) -- (2,2);
\draw[green2,very thick, dashed] (0.280961, -1.38602) -- (0.561922, -2.77205);
\draw[green2,very thick, dashed] (0.67801, -1.24109) -- (1.35602, -2.48218);
\draw[blue, very thick, dashed] (-1.4,0.4) arc (90:286:0.5 and 0.5) ;
\draw[blue, very thick, dashed] (0.1,2.7) arc (197:317:0.8 and 0.8) ;
\draw[blue, very thick, decoration={snake,amplitude=2pt, segment length=4pt}, decorate]
(-0.4,2.76) arc (197:315:1.3 and 1.3) ;
\draw[blue, very thick, decoration={snake,amplitude=2pt, segment length=4pt}, decorate]
(1.25,-0.75) arc (-90:90:0.7 and 0.7) ;
\draw[blue, very thick, decoration={snake,amplitude=2pt, segment length=4pt}, decorate]
(-0.,-2.76) arc (0:158:0.7 and 0.7) ;
\draw[blue, very thick, dashed] (2.5,-1.3) arc (-95:-300:0.5 and 0.5) ;
\draw[blue, very thick, dashed] (2.5, 1.3) arc (95:300:0.5 and 0.5) ;
\draw[very thick] (0,0) circle (40pt);
\draw[very thick] (0,0) circle (80pt);
\end{tikzpicture}
\caption{\label{ldd}\small Ladder (green) and rainbow (blue) propagators.}
\end{center}
\end{figure}
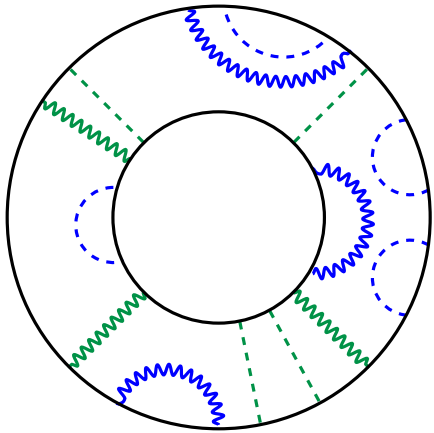

Diagrams that survive are constructed from two  building blocks (fig.~\ref{ldd}): ladder propagators that connect different loops and  rainbow propagators attached to the same loop. These two elements are in a way similar to the worldsheets of different topology: ladders correspond to a cylinder worldsheet that connects a pair of Wilson loops, while rainbow diagrams correspond to a disk attached to a single contour.
This  analogy is rather loose so long as a single diagram is concerned, because a generic diagram will contain both types of propagators in equal proportion.

Similarity to string theory becomes more pronounced at strong coupling when propagators tend to become dense. Indeed the leading, dominant contribution then comes from diagrams of order\footnote{This counting  follows from the area law behavior at strong coupling, and is shared by the ladder approximation. The argument is rather simple and is outlined in the appendix~\ref{average-l}.} $\ell\sim \mathcal{O}(\sqrt{\lambda })$.
Depending on the parameters of the problem, only one type of propagators will appear with $\mathcal{O}(\sqrt{\lambda })$ multiplicity, while the number of propagators of the other type will be much smaller, $\mathcal{O}(1)$. As a result, the leading diagrams at strong coupling are almost exclusively built either from ladder or from rainbow propagators. The competition between the two contributions leads to a phase transition \cite{Zarembo:2001jp}, analogous to the Gross-Ooguri transition in string theory which is caused by competition between connected and disconnected minimal surfaces.

In the ladder approximation the problem becomes effectively one-dimen\-sional, because the 4d fields only appear in the combinations
\begin{equation}
{\cal O}_a (t)  = i A_\mu \dot x^\mu_a +\Phi_I  n^I_a |\dot x_a|,
\end{equation}
defined on each loop in the correlator. The fields $\mathcal{O}_a(t)$ are linear in $A_\mu $ and $\Phi _I$ and thus are Gaussian with the effective propagators
\begin{equation}
\begin{split}
 \left\langle {\cal O}^i_{a\,j}(t)\bar{{\cal O}}_{b\,l}^k(s)\right\rangle &=
 \frac{1}{N}\,\delta ^i_l\delta ^k_jG_{ab}(t-s),
 \\
 \left\langle {\cal O}_{a\,j}^i(t){\cal O}_{b\,l}^k(s)\right\rangle &=
 \frac{1}{N}\,\delta ^i_l\delta ^k_j\widetilde G_{ab}(t-s),
\end{split}
\end{equation}
where $i\ldots l$ are the color indices and the bar again corresponds to a contour of the opposite orientation.

The propagator connecting two points on the same circle is a constant:
\begin{equation}\label{Gaa}
 \widetilde{G}_{aa}=\frac{\lambda }{16\pi ^2}\equiv g,
\end{equation}
while for different circles the propagators become
\begin{equation}\label{G(theta)}
G_{ab}(\theta )=\frac{\lambda }{16\pi ^2}\,\,\frac{\cos\gamma_{ab} +\cos\theta }{\frac{R_a^2+R_b^2+h_{ab}^2}{2R_aR_b}-\cos\theta }\equiv G(\theta ),
\end{equation}
\begin{equation}\label{G(thetatilde)}
\widetilde G_{ab}(\theta )= \frac{\lambda }{16\pi ^2}\,\frac{\cos\gamma_{ab} -\cos\theta }{\frac{R_a^2+R_b^2+h_{ab}^2}{2R_aR_b}-\cos\theta } \equiv  \widetilde G(\theta ),
\end{equation}
where $\gamma_{ab}$ and $h_{ab}$ stand for the differences $\gamma_a-\gamma_b$ and $h_a-h_b$. It is easy to see that (\ref{G(thetatilde)}) reduces to (\ref{Gaa}) for $R_a=R_b$, $h_{ab}=0$, and $\gamma _{ab}=0$.

We start by considering the connected correlator of two loops with opposite orientations:
\begin{equation}
  \left\langle W(C_1)W(\bar C_2)\right\rangle_{\rm conn} =
 \left\langle W(C_1)W(\bar C_2)\right\rangle
 -\left\langle W(C_1)\rangle\langle W(\bar C_2)\right\rangle.
\end{equation}
As in (\ref{WLdef}), the Wilson loops  can be defined by the path-ordered exponentials:
\begin{equation}
 \PU_a(t_1,t_2)=\overrightarrow{P}\exp\int_{t_1}^{t_2}dt\, {\cal O} _a(t),
 \qquad
  \AU_a(t_1,t_2)=\overleftarrow{P}\exp\int_{t_1}^{t_2}dt\,{\cal O}_a(t),
\label{PUAU}
\end{equation}
where $\overrightarrow{P}$ and $\overleftarrow{P}$ denote path and anti-path ordering. The closed contour corresponds to $t_{1}=0$ and $t_{2}=2\pi $, but for the sake of deriving a complete set of Dyson equations we will need to consider an arc line between generic $t_{1}$ and $t_{2}$.

In the ladder approximation,
\begin{equation}
 \left\langle W(C_1)W(\bar C_2)\right\rangle_{\rm conn}\stackrel{\rm ladd.}{=}
  \langle
 \mathop{\mathrm{tr}}\AU_1(0,2\pi )
 \mathop{\mathrm{tr}}\PU_2(0,2\pi )
 \rangle_{\rm conn},
\label{def:corr}
\end{equation}
where the bracket on the right-hand-side denotes Gaussian average defined by the propagators (\ref{Gaa}), (\ref{G(theta)}).

The key technical simplification of the ladder approximation is that the diagrams that survive can be generated by iterating certain integral equations. These equations can then be used for analytic diagram resummation. To derive a closed set of Dyson equations we need Green's functions of two types:
\begin{equation}
 K_{ab}(t)= \langle \mathop{\mathrm{tr}}\AU_a(0,t)\mathop{\mathrm{tr}}\PU_b(0,2\pi )\rangle_{\rm conn}
\label{K}
\end{equation}
\begin{equation}\label{Gamma}
 \Gamma_{ab} (t,s|\varphi )= \frac{1}{N}\langle\mathop{\mathrm{tr}}\AU_a(0,t)\PU_b(\varphi ,\varphi +s)\rangle.
\end{equation}
The Wilson loop correlator is expressed through $K_{12}$ evaluated at $t=2\pi $:
\begin{equation}
 \left\langle W(C_1)W(\bar C_2)\right\rangle_{\rm conn}\stackrel{\rm ladd.}
 =K_{12}(2\pi ),
\end{equation}
while $\Gamma _{{ab}}$ plays an auxiliary role.

The Dyson equation that relates $K_{ab}$ to $\Gamma_{ab}$ is derived in the appendix~\ref{dyson-appendix}:
\begin{equation}\label{DysonK}
 K_{ab}(t)= 2g\!\int_{0}^{t}\!dt'\!\int_{0}^{t'}\!dt''\,W(t'-t'')K_{ab}(t'')
 +\int_{0}^{t}\!dt'\!\int_{0}^{2\pi }\!d\varphi \,G(\varphi -t')\Gamma_{ab}(t',2\pi |\varphi ).
\end{equation}
where
\begin{equation}\label{defW0}
 W(t) =\frac{1}{N} \langle\mathop{\mathrm{tr}}\AU_a(0,t)\rangle,
\end{equation}
This relation is similar to the Dyson equation in \cite{Zarembo:2001jp}, but is not exactly equivalent to it. We have checked that the new equation correctly reproduces combinatorics of ladder diagrams for the super\-symmetric configuration of Wilson loops considered in \cite{Correa:2018lyl}.

\begin{figure}[h]
\begin{tikzpicture}[scale=0.87]
\draw[thick,black!35] (1.5,2) -- (2.5,1);
\draw[thick,black!35] (3.4,2) -- (2.2,1);
\draw[thick,black!35] (1.5,0) -- (2.5,1);
\draw[thick,black!35] (3.1,0) -- (2.,1);
\draw[black!35] (2.35,1.7) node {$\cdots$};
\draw[black!35] (2.3,0.25) node {$\cdots$};
\draw[thick,dashed] (1,2) .. controls (1.1,2.55) and (3.9,2.55) .. (4,2);
\draw[thick,red,->] (1,2) -- (2.5,2);
\fill[red] (1,2) circle (1pt);
\draw[thick,red] (2.4,2) -- (4,2);
\fill[red] (4,2) circle (1pt);
\draw[thick,dashed] (1,0) .. controls (1.1,-.55) and (3.3,-.55) .. (3.4,0);
\draw[thick,red,] (1,0) -- (2.3,0);
\fill[red] (1,0) circle (1pt);
\draw[thick,red,<-] (2.2,0) -- (3.4,0);
\fill[red] (3.4,0) circle (1pt);

\fill[oran] (2.3,1) ellipse (17pt and 11pt);
\draw[black,thick] (2.3,1) ellipse (17pt and 11pt);

\draw[font=\scriptsize] (1.0,1.75) node {$0$};
\draw[font=\scriptsize] (4,1.75) node {$2\pi$};
\draw[font=\scriptsize] (1,0.2) node {$0$};
\draw[font=\scriptsize] (3.4,0.2) node {$t$};

\draw (4.6,1) node  {$=K(t)$};
\end{tikzpicture}
\hspace{0.2cm}
\begin{tikzpicture}[scale=0.87,baseline={(0,-0.53)}]
\draw[thick,black!35] (1.7,2) -- (2.5,1);
\draw[thick,black!35] (3.2,2) -- (2.2,1);
\draw[thick,black!35] (1.5,0) -- (2.5,1);
\draw[thick,black!35] (3.1,0) -- (2.,1);
\draw[black!35] (2.35,1.7) node {$\cdots$};
\draw[black!35] (2.3,0.25) node {$\cdots$};

\draw[thick,dashed] (1,0) -- (1.3,2);

\draw[thick,red,->] (1.3,2) -- (2.7,2);
\fill[red] (1.3,2) circle (1pt);
\draw[thick,red] (2.6,2) -- (3.8,2);
\fill[red] (3.8,2) circle (1pt);
\draw[thick,red,] (1,0) -- (2.3,0);
\fill[red] (1,0) circle (1pt);
\draw[thick,red,<-] (2.2,0) -- (3.4,0);
\fill[red] (3.4,0) circle (1pt);

\fill[pattern=dots, magenta2] (2.3,1) ellipse (17pt and 11pt);
\draw[black,thick] (2.3,1) ellipse (17pt and 11pt);

\draw[font=\scriptsize] (1.3,2.15) node {$\varphi$};
\draw[font=\scriptsize] (3.8,2.15) node {$\varphi+s$};
\draw[font=\scriptsize] (1,-0.2) node {$0$};
\draw[font=\scriptsize] (3.4,-0.2) node {$t$};

\draw[font=\scriptsize] (3.6,0.1) node {$i$};
\draw[font=\scriptsize] (4,1.9) node {$j$};

\draw (4.9,1) node  {$=\Gamma(t,s|\varphi)\delta^i_j$};
\end{tikzpicture}
\hspace{-0.2cm}
\begin{tikzpicture}[scale=0.87,baseline={(-0.7,-0.53)}]
\draw[thick,black!35] (1.5,0) -- (2.5,1);
\draw[thick,black!35] (3.1,0) -- (2.,1);
\draw[black!35] (2.3,0.25) node {$\cdots$};

\draw[thick,red,] (1,0) -- (2.3,0);
\fill[red] (1,0) circle (1pt);
\draw[thick,red,<-] (2.2,0) -- (3.4,0);
\fill[red] (3.4,0) circle (1pt);

\fill[cyan2] (2.3,1) ellipse (17pt and 11pt);
\draw[black,thick] (2.3,1) ellipse (17pt and 11pt);

\draw[font=\scriptsize] (1,-0.2) node {$0$};
\draw[font=\scriptsize] (3.4,-0.2) node {$t$};

\draw[font=\scriptsize] (3.6,0) node {$i$};
\draw[font=\scriptsize] (0.8,0) node {$j$};
\draw (4.2,1.) node  {$=W(t)\delta^i_j$};
\end{tikzpicture}
\caption{Diagrammatic representation of Green's functions}
\label{greenrep}
\end{figure}
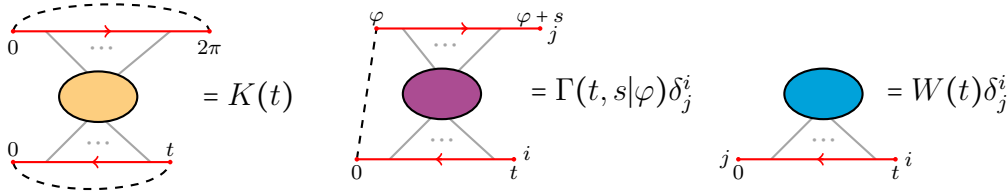

In order to better understand eq. (\ref{DysonK}) diagrammatically, we represent the Green's functions (\ref{K})-(\ref{Gamma}),  as well as (\ref{defW0}), as shown in figure \ref{greenrep}.

Propagators, represented by blue and green dashed double lines, can be of two sorts depending on whether they connect two points in the same or different loops:
\begin{figure}[h]
\center{
\begin{tikzpicture}[scale=1]
\draw[thick,blue,dashed] (1,0.1) -- (3,0.1);
\draw[thick,blue,dashed] (1,-0.1) -- (3,-0.1);
\draw (4.3,0) node  {$=\frac{g}{N}\delta^i_j \delta^k_l$};
\draw (0.8,0.2) node  {$l$};
\draw (0.8,-0.2) node  {$i$};
\draw (3.2,0.2) node  {$k$};
\draw (3.2,-0.2) node  {$j$};
\end{tikzpicture}
\hspace{0.5cm}
\begin{tikzpicture}[scale=1]
\draw[thick,green2,dashed] (1,0.1) -- (3,0.1);
\draw[thick,green2,dashed] (1,-0.1) -- (3,-0.1);
\draw (4.5,0) node  {$=\frac{G(\theta)}{N}\delta^i_j \delta^k_l$};
\draw (0.8,0.2) node  {$l$};
\draw (0.8,-0.2) node  {$i$};
\draw (3.2,0.2) node  {$k$};
\draw (3.2,-0.2) node  {$j$};
\end{tikzpicture}}
\end{figure}

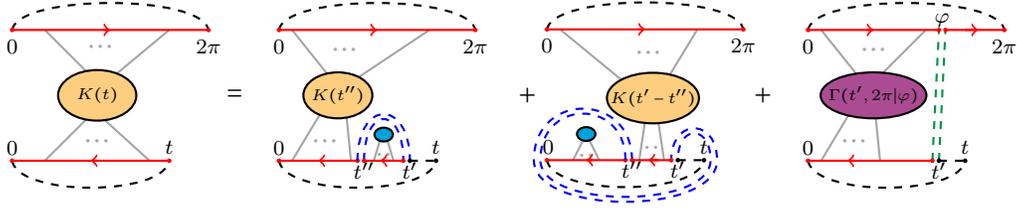
\begin{figure}[h]
\begin{tikzpicture}[scale=0.87]

\draw[thick,black!35] (1.5,2) -- (2.5,1);
\draw[thick,black!35] (3.4,2) -- (2.2,1);
\draw[thick,black!35] (1.5,0) -- (2.5,1);
\draw[thick,black!35] (3.1,0) -- (2.,1);
\draw[black!35] (2.35,1.7) node {$\cdots$};
\draw[black!35] (2.3,0.25) node {$\cdots$};

\draw[thick,dashed] (1,2) .. controls (1.1,2.55) and (3.9,2.55) .. (4,2);
\draw[thick,red,->] (1,2) -- (2.5,2);
\fill[red] (1,2) circle (1pt);
\draw[thick,red] (2.4,2) -- (4,2);
\fill[red] (4,2) circle (1pt);
\draw[thick,dashed] (1,0) .. controls (1.1,-.55) and (3.3,-.55) .. (3.4,0);
\draw[thick,red,] (1,0) -- (2.3,0);
\fill[red] (1,0) circle (1pt);
\draw[thick,red,<-] (2.2,0) -- (3.4,0);
\fill[red] (3.4,0) circle (1pt);

\fill[oran] (2.3,1) ellipse (17pt and 11pt);
\draw[black,thick] (2.3,1) ellipse (17pt and 11pt);
\draw[font=\tiny] (2.3,1) node  {$K(t)$};
\draw[font=\scriptsize] (1.0,1.75) node {$0$};
\draw[font=\scriptsize] (4,1.75) node {$2\pi$};
\draw[font=\scriptsize] (1,0.2) node {$0$};
\draw[font=\scriptsize] (3.4,0.2) node {$t$};

\draw (4.4,1) node {$=$};
\end{tikzpicture}
\begin{tikzpicture}[scale=0.87]

\draw[thick,black!35] (1.2,2) -- (2.,1);
\draw[thick,black!35] (3.3,2) -- (1.8,1);
\draw[thick,black!35] (1.2,0) -- (1.8,1);
\draw[thick,black!35] (2.1,0) -- (2,1);
\draw[black!35] (2.,1.65) node {$\cdots$};
\draw[black!35] (1.7,0.25) node {$\cdots$};

\draw[thick,black!35] (2.45,0) -- (2.6,0.5);
\draw[thick,black!35] (2.75,0) -- (2.6,0.5);
\draw[black!35] (2.6,0.1) node {..};

\draw[thick,dashed] (1,2) .. controls (1.1,2.55) and (3.9,2.55) .. (4,2);
\draw[thick,red,->] (1,2) -- (2.5,2);
\fill[red] (1,2) circle (1pt);
\draw[thick,red] (2.4,2) -- (4,2);
\fill[red] (4,2) circle (1pt);
\draw[thick,dashed] (1,0) .. controls (1.1,-.55) and (3.3,-.55) .. (3.4,0);
\draw[thick,red] (1,0) -- (1.7,0);
\fill[red] (1,0) circle (1pt);
\draw[thick,red,<-] (1.6,0) -- (2.2,0);
\fill[red] (2.2,0) circle (1pt);

\draw[thick,dashed,blue] (2.2,0) .. controls (2.2,.95) and (3.,.95) .. (3,0);
\draw[thick,dashed,blue] (2.3,0) .. controls (2.3,.85) and (2.9,.85) .. (2.9,0);

\draw[thick,red] (2.3,0) -- (2.7,0);
\fill[red] (2.3,0) circle (1pt);
\draw[thick,red,<-] (2.55,0) -- (2.9,0);
\fill[red] (2.9,0) circle (1pt);

\fill[black] (3,0) circle (1pt);
\draw[thick,dashed] (3,0) -- (3.4,0);
\fill[black] (3.4,0) circle (1pt);

\fill[oran] (1.9,1) ellipse (15pt and 10pt);
\draw[black,thick] (1.9,1) ellipse (15pt and 10pt);
\draw[font=\tiny] (1.9,1) node  {$K(t'')$};

\draw[font=\scriptsize] (1,1.75) node {$0$};
\draw[font=\scriptsize] (4,1.75) node {$2\pi$};
\draw[font=\scriptsize] (1,0.2) node {$0$};
\draw[font=\scriptsize] (3.4,0.2) node {$t$};
\draw[font=\scriptsize] (3.,-0.15) node {$t'$};
\draw[font=\scriptsize] (2.3,-0.15) node {$t''$};

\fill[cyan2] (2.6,0.4) ellipse (4pt and 3pt);
\draw[black,thick] (2.6,0.4) ellipse (4pt and 3pt);

\end{tikzpicture}
\begin{tikzpicture}[scale=0.87,baseline={(0,-0.5)}]

\draw (0.7,1) node {$+$};

\draw[thick,black!35] (1.2,2) -- (2.6,0.85);
\draw[thick,black!35] (3.3,2) -- (2.6,0.85);
\draw[thick,black!35] (1.4,0) -- (1.6,0.4);
\draw[thick,black!35] (1.8,0) -- (1.6,0.4);
\draw[black!35] (2.55,1.6) node {$\cdots$};
\draw[black!35] (1.6,0.075) node {..};

\draw[thick,black!35] (2.41,0) -- (2.5,0.6);
\draw[thick,black!35] (2.79,0) -- (2.7,0.6);
\draw[black!35] (2.6,0.2) node {..};

\draw[thick,dashed] (1,2) .. controls (1.1,2.55) and (3.9,2.55) .. (4,2);
\draw[thick,red,->] (1,2) -- (2.5,2);
\fill[red] (1,2) circle (1pt);
\draw[thick,red] (2.4,2) -- (4,2);
\fill[red] (4,2) circle (1pt);
\draw[thick,dashed] (1,0) .. controls (1.1,-.55) and (3.3,-.55) .. (3.4,0);
\draw[thick,red] (1,0) -- (1.7,0);
\fill[red] (1,0) circle (1pt);
\draw[thick,red,<-] (1.6,0) -- (2.2,0);
\fill[red] (2.2,0) circle (1pt);

\draw[thick,dashed,blue]
 (2.2,0) .. controls (2.2,.95) and (0.9,.95)..
 (0.9,0) .. controls (0.9,-0.75) and (3.5,-0.75)..
 (3.5,0) .. controls (3.5,0.55) and (3.,0.55)
.. (3,0);

\draw[thick,dashed,blue]
 (2.3,0) .. controls (2.3,1.05) and (0.8,1.05)..
 (0.8,0) .. controls (0.8,-0.85) and (3.6,-0.85)..
 (3.6,0) .. controls (3.6,0.65) and (2.9,0.65)
.. (2.9,0);

\draw[thick,red] (2.3,0) -- (2.7,0);
\fill[red] (2.3,0) circle (1pt);
\draw[thick,red,<-] (2.55,0) -- (2.9,0);
\fill[red] (2.9,0) circle (1pt);

\fill[black] (3,0) circle (1pt);
\draw[thick,dashed] (3,0) -- (3.4,0);
\fill[black] (3.4,0) circle (1pt);

\fill[oran] (2.62,0.95) ellipse (20pt and 11pt);
\draw[black,thick] (2.62,0.95) ellipse (20pt and 11pt);
\draw[font=\tiny] (2.62,0.95) node  {$K(t'-t'')$};

\draw[font=\scriptsize] (1,1.75) node {$0$};
\draw[font=\scriptsize] (4,1.75) node {$2\pi$};
\draw[font=\scriptsize] (1.02,0.2) node {$0$};
\draw[font=\scriptsize] (3.4,0.2) node {$t$};
\draw[font=\scriptsize] (3.,-0.15) node {$t'$};
\draw[font=\scriptsize] (2.3,-0.15) node {$t''$};

\fill[cyan2] (1.6,0.4) ellipse (4pt and 3pt);
\draw[black,thick] (1.6,0.4) ellipse (4pt and 3pt);

\draw (4.3,1) node {$+$};
\end{tikzpicture}
\begin{tikzpicture}[scale=0.87]

\draw[thick,dashed,green2] (3,0) -- (3.1,2);
\draw[thick,dashed,green2] (2.9,0)--(3,2);

\draw[thick,black!35] (1.2,2) -- (2.,1);
\draw[thick,black!35] (2.8,2) -- (1.8,1);
\draw[thick,black!35] (1.2,0) -- (1.8,1);
\draw[thick,black!35] (2.1,0) -- (2,1);
\draw[black!35] (2.,1.65) node {$\cdots$};
\draw[black!35] (1.7,0.25) node {$\cdots$};

\draw[thick,dashed] (1,2) .. controls (1.1,2.55) and (3.9,2.55) .. (4,2);
\draw[thick,red,->] (1,2) -- (2.,2);
\fill[red] (1,2) circle (1pt);
\draw[thick,red] (1.9,2) -- (3,2);
\fill[red] (3,2) circle (1pt);

\draw[thick,red,->] (3.1,2) -- (3.6,2);
\fill[red] (3.1,2) circle (1pt);
\draw[thick,red] (3.4,2) -- (4,2);
\fill[red] (4,2) circle (1pt);

\draw[thick,dashed] (1,0) .. controls (1.1,-.55) and (3.3,-.55) .. (3.4,0);

\draw[thick,red] (1,0) -- (1.7,0);
\fill[red] (1,0) circle (1pt);
\draw[thick,red,<-] (1.6,0) -- (2.9,0);
\fill[red] (2.9,0) circle (1pt);

\fill[black] (3,0) circle (1pt);
\draw[thick,dashed] (3,0) -- (3.4,0);
\fill[black] (3.4,0) circle (1pt);

\fill[magenta2] (2,1) ellipse (23pt and 10pt);
\draw[black,thick] (2,1) ellipse (23pt and 10pt);
\draw[font=\tiny] (2,1) node  {$\Gamma(t',2\pi|\varphi)$};

\draw[font=\scriptsize] (1,1.75) node {$0$};
\draw[font=\scriptsize] (4,1.75) node {$2\pi$};
\draw[font=\scriptsize] (1,0.2) node {$0$};
\draw[font=\scriptsize] (3.4,0.2) node {$t$};
\draw[font=\scriptsize] (3.,-0.15) node {$t'$};
\draw[font=\scriptsize] (3.05,2.15) node {$\varphi$};

\end{tikzpicture}
\caption{Diagrammatic interpretation of the integral equation (\ref{DysonK})}
\label{Kcorre1}
\end{figure}

In eq. (\ref{DysonK}) $t'$ indicates the position of the rightmost field in $\AU_a(0,t)$ contracted with a propagator. This contraction could be either with another field in $\AU_a(0,t)$ sitting at a point $t''<t'$ or with a field in $\PU_b(0,2\pi)$ sitting at a point $\varphi$. In the former case, there are two planar contributions, depicted by the first two diagrams on the right-hand-side of the equation shown in figure \ref{Kcorre1}, but those contributions are equivalent upon a change of integration variables. For the latter case, we get the last diagram in figure \ref{Kcorre1}, which corresponds to the last term in the right-hand-side of eq. (\ref{DysonK}).

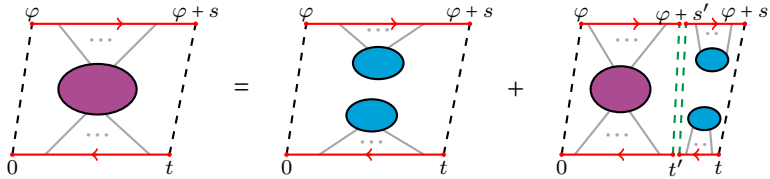
\begin{figure}[h]
\center{
\begin{tikzpicture}[scale=0.87,baseline={(0,-0.53)}]
\draw[thick,black!35] (1.7,2) -- (2.5,1);
\draw[thick,black!35] (3.2,2) -- (2.2,1);
\draw[thick,black!35] (1.5,0) -- (2.5,1);
\draw[thick,black!35] (3.1,0) -- (2.,1);
\draw[black!35] (2.35,1.7) node {$\cdots$};
\draw[black!35] (2.3,0.25) node {$\cdots$};

\draw[thick,dashed] (1,0) -- (1.3,2);
\draw[thick,dashed] (3.4,0) -- (3.8,2);

\draw[thick,red,->] (1.3,2) -- (2.7,2);
\fill[red] (1.3,2) circle (1pt);
\draw[thick,red] (2.6,2) -- (3.8,2);
\fill[red] (3.8,2) circle (1pt);
\draw[thick,red,] (1,0) -- (2.3,0);
\fill[red] (1,0) circle (1pt);
\draw[thick,red,<-] (2.2,0) -- (3.4,0);
\fill[red] (3.4,0) circle (1pt);

\fill[magenta2] (2.3,1) ellipse (17pt and 11pt);
\draw[black,thick] (2.3,1) ellipse (17pt and 11pt);

\draw[font=\scriptsize] (1.3,2.15) node {$\varphi$};
\draw[font=\scriptsize] (3.8,2.15) node {$\varphi+s$};
\draw[font=\scriptsize] (1,-0.2) node {$0$};
\draw[font=\scriptsize] (3.4,-0.2) node {$t$};

\draw (4.5,1) node {$=$};

\end{tikzpicture}
\begin{tikzpicture}[scale=0.87,baseline={(0,-0.53)}]
\draw[thick,black!35] (1.8,2) -- (2.5,1.4);
\draw[thick,black!35] (3.1,2) -- (2.2,1.4);
\draw[thick,black!35] (1.5,0) -- (2.4,0.6);
\draw[thick,black!35] (3.1,0) -- (2.1,0.6);
\draw[black!35] (2.4,1.85) node {$\cdots$};
\draw[black!35] (2.3,0.15) node {$\cdots$};

\draw[thick,dashed] (1,0) -- (1.3,2);
\draw[thick,dashed] (3.4,0) -- (3.8,2);

\draw[thick,red,->] (1.3,2) -- (2.7,2);
\fill[red] (1.3,2) circle (1pt);
\draw[thick,red] (2.6,2) -- (3.8,2);
\fill[red] (3.8,2) circle (1pt);
\draw[thick,red,] (1,0) -- (2.3,0);
\fill[red] (1,0) circle (1pt);
\draw[thick,red,<-] (2.2,0) -- (3.4,0);
\fill[red] (3.4,0) circle (1pt);

\fill[cyan2] (2.4,1.4) ellipse (11pt and 7pt);
\draw[black,thick] (2.4,1.4) ellipse (11pt and 7pt);

\fill[cyan2] (2.3,0.6) ellipse (11pt and 7pt);
\draw[black,thick] (2.3,0.6) ellipse (11pt and 7pt);

\draw[font=\scriptsize] (1.3,2.15) node {$\varphi$};
\draw[font=\scriptsize] (3.8,2.15) node {$\varphi+s$};
\draw[font=\scriptsize] (1,-0.2) node {$0$};
\draw[font=\scriptsize] (3.4,-0.2) node {$t$};

\draw (4.5,1) node {$+$};
\end{tikzpicture}
\begin{tikzpicture}[scale=0.87,baseline={(0,-0.53)}]
\draw[thick,black!35] (1.4,2) -- (2.,1);
\draw[thick,black!35] (2.6,2) -- (1.8,1);
\draw[thick,black!35] (1.2,0) -- (2.,1);
\draw[thick,black!35] (2.5,0) -- (1.8,1);
\draw[black!35] (2.,1.7) node {$\cdots$};
\draw[black!35] (1.85,0.25) node {$\cdots$};

\draw[thick,black!35] (3.05,2) -- (3.15,1.4);
\draw[thick,black!35] (3.6,2) -- (3.5,1.4);
\draw[thick,black!35] (2.9,0) -- (3.05,0.6);
\draw[thick,black!35] (3.3,0) -- (3.25,0.6);
\draw[black!35] (3.3,1.85) node {$..$};
\draw[black!35] (3.12,0.15) node {$..$};

\fill[cyan2] (3.3,1.45) ellipse (7pt and 5pt);
\draw[black,thick] (3.3,1.45) ellipse (7pt and 5pt);

\fill[cyan2] (3.18,0.55) ellipse (7pt and 5pt);
\draw[black,thick] (3.18,0.55) ellipse (7pt and 5pt);

\draw[thick,dashed,green2] (2.7,0) -- (2.8,2);
\draw[thick,dashed,green2] (2.8,0) -- (2.9,2);

\draw[thick,dashed] (1,0) -- (1.3,2);
\draw[thick,dashed] (3.4,0) -- (3.8,2);

\draw[thick,red,->] (1.3,2) -- (2.2,2);
\fill[red] (1.3,2) circle (1pt);

\draw[thick,red] (2.1,2) -- (2.8,2);
\fill[red] (2.8,2) circle (1pt);

\draw[thick,red,->] (2.9,2) -- (3.4,2);
\fill[red] (2.9,2) circle (1pt);

\draw[thick,red] (3.3,2) -- (3.8,2);
\fill[red] (3.8,2) circle (1pt);

\draw[thick,red,] (1,0) -- (1.9,0);
\fill[red] (1,0) circle (1pt);
\draw[thick,red,<-] (1.85,0) -- (2.7,0);
\fill[red] (2.7,0) circle (1pt);

\draw[thick,red] (2.8,0) -- (3.1,0);
\fill[red] (2.8,0) circle (1pt);

\draw[thick,red,<-] (3.,0) -- (3.4,0);
\fill[red] (3.4,0) circle (1pt);

\fill[magenta2] (1.9,1) ellipse (13pt and 10pt);
\draw[black,thick] (1.9,1) ellipse (13pt and 10pt);

\draw[font=\scriptsize] (1.3,2.15) node {$\varphi$};
\draw[font=\scriptsize] (3.8,2.15) node {$\varphi+s$};
\draw[font=\scriptsize] (1,-0.2) node {$0$};
\draw[font=\scriptsize] (3.4,-0.2) node {$t$};

\draw[font=\scriptsize] (2.85,2.15) node {$\varphi+s'$};
\draw[font=\scriptsize] (2.75,-0.2) node {$t'$};

\end{tikzpicture}
\caption{Diagrammatic interpretation of the integral equation (\ref{improvedDysonG})}
\label{Gcorre1}}
\end{figure}

The Dyson equation for the auxiliary Green's function $\Gamma_{ab}(t,s|\varphi )$ closes on itself:
\begin{equation}\label{improvedDysonG}
 \Gamma_{ab}(t,s|\varphi )=W(t)W(s)+\int_{0}^{t}\!\!dt'\!\!\int_{0}^{s}\!\!ds'\,
 W(t-t')W(s-s')G(\varphi +s'-t')\Gamma_{ab}(t',s'|\varphi ).
\end{equation}
An analytic derivation is presented in the appendix~\ref{dyson-appendix}. Diagrammatically, the Dyson equation can also be understood as follows. The first term comes from diagrams with no propagator connecting the two loops. In the second term $t'$ stands for the rightmost point in $\AU_a(0,t)$ with a propagator connecting with a point $\varphi+s'$ in $\PU_b(\varphi,\varphi+s)$, as shown in figure \ref{Gcorre1}. Thus, in the planar limit, to the right of $t'$ we can only have propagators within the segment $(t',t)$ and similarly,  to the right of $\varphi+s'$ we can only have propagators within the segment $(\varphi+s',\varphi+s)$.

The same analysis can be repeated for two loops with the same orientation, in which case the Green's functions are defined as
\begin{equation}
	\widetilde K_{ab}(t)= \langle \mathop{\mathrm{tr}}\PU_a(0,t)\mathop{\mathrm{tr}}\PU_b(0,2\pi )\rangle_{\rm conn},
	\label{Ktilde}
\end{equation}
\begin{equation}\label{Gammatilde}
	\widetilde \Gamma_{ab} (t,s|\varphi )= \frac{1}{N}\langle\mathop{\mathrm{tr}}\PU_a(0,t)\PU_b(\varphi-s,\varphi)\rangle.
	\end{equation}
An equation that relates the two functions is essentially equivalent to (\ref{DysonK}):
\begin{equation}\label{DysonKtilde}
	\widetilde K_{ab}(t)= 2g\!\int_{0}^{t}\!dt'\!\int_{0}^{t'}\!dt''\,W(t'-t'')\widetilde K_{ab}(t'')
	+\int_{0}^{t}\!dt'\!\int_{0}^{2\pi }\!d\varphi \,\widetilde G(\varphi -t')\widetilde\Gamma_{ab}(t',2\pi |\varphi ),
\end{equation}
while the auxiliary Dyson equation is slightly different:
\begin{equation}\label{improvedDysonGtilde}
 \widetilde\Gamma_{ab}(t,s|\varphi )=W(t)W(s)+\int_{0}^{t}\!\!dt'\!\!\int_{0}^{s}\!\!ds'\,
 W(t-t')W(s-s')\widetilde G(\varphi -s'-t')\widetilde\Gamma_{ab}(t',s'|\varphi ),
\end{equation}
reflecting the fact that the endpoints of the ladder propagators for parallel circles must be arranged in a different order compared to the case of contours of opposite orientation.

\section{Solving Dyson equations}

We first consider the connected correlator of two Wilson loops of opposite orientation.
To account for the ladder contribution we need to solve (\ref{improvedDysonG}) and then express $K_{ab}$ in terms of $\Gamma_{ab}$ using (\ref{DysonK}).
We start with the latter step.

The Dyson equation (\ref{DysonK}) has the following form:
\begin{equation}
 f(t)=2g\int_{0}^{t}dt'\int_{0}^{t'}dt''\,W(t'-t'')f (t'')+\int_{0}^{t}dt'\,j(t'),
 \label{ineq2g}
\end{equation}
This is an integral equation of convolution type and, as such,  can be solved by the Laplace transform:
\begin{equation}
 f(z)=\int_{0}^{\infty }dt\,\,{\rm e}\,^{-zt}f(t).
\end{equation}
Taking into account that the Laplace image of $W(t)$ is\footnote{See appendix~\ref{dyson-appendix} or  \ref{MatrixModelResults}.}
\begin{equation}
 W(z)=\frac{z-\sqrt{z^{2}-4g}}{2g}\,,
\end{equation}
solving for $f(z)$, and going back to the original variables we find:
\begin{equation}
 f(t)=\int\limits_{0}^{t}dt'\,V(t-t')j(t'),
\end{equation}
where the kernel is given by
\begin{equation}
V(z)=\frac{1}{\sqrt{z^2-4g}}\qquad \Longrightarrow\qquad V(t) =  I_0(2\sqrt{g}t).
\end{equation}

Applying this result to (\ref{DysonK}) we get
\begin{equation}
 K_{ab}(t)=\int_{0}^{t}dt'\,\int_{0}^{2\pi }d\varphi \,V(t-t')
 G(\varphi -t')\Gamma_{ab} (t',2\pi |\varphi ).
\label{kW1W2}
\end{equation}
The ladder contribution to the connected correlator of two loops with opposite  orientations is obtained from this equation
as $K_{12}(2\pi)$.

Similarly, the ladder contribution in the case of loops with the same orientations can be worked out from
\begin{equation}
 \widetilde K_{ab}(t)=\int_{0}^{t}dt'\,\int_{0}^{2\pi }d\varphi \,V(t-t')
 \widetilde G(\varphi -t')\widetilde\Gamma_{ab} (t',2\pi |\varphi ).
\label{ktildeW1W2}
\end{equation}

Thus, in  order to have explicit expressions for the ladder contribution to correlators of two loops it is sufficient to solve the integral equations for the auxiliary Green's functions (\ref{improvedDysonG}) and  (\ref{improvedDysonGtilde}).
As we shall see the problem reduces to a one-dimensional Schr\"odinger equation for a particle in a periodic potential, which will allow us to obtain a spectral representation for the correlator. In a special case when $h_{ab}$ and $\gamma_{ab}$ are related such as to render the effective propagator $G$ constant, the solution can be found explicitly at any coupling. The spectral representation also considerably simplifies in the strong-coupling limit.

\subsection{Spectral representation for opposite orientations}

As shown in \cite{Zarembo:2001jp}, the solution of the Dyson equation (\ref{improvedDysonG}) admits a spectral representation in terms of the eigenfunctions of a certain Schr\"odinger operator. The Schr\"odinger representation arises upon changing variables to
\begin{equation}
 x=s-t,\qquad y=s+t.
 \label{change}
\end{equation}
We use the same notation $\Gamma(x,y|\varphi ) $ for the Green's function in the new variables, which hopefully will not cause any confusion\footnote{And also omit the indices $ab$ labeling the loops.}. While the function $\Gamma (t,s|\varphi )$ is defined in the upper right quadrant of the $(s,t)$ plane, the new variables span a wedge $y>|x|$. The kernel  $\Gamma(x,y|\varphi ) $ is an exponentially growing function of $y$, for any fixed $x$, satisfying boundary condition $\Gamma (x,|x||\varphi )=W(|x|)$. It is natural, therefore, to Laplace transform in $y$:
\begin{equation}
 \Gamma (x,y|\varphi )\rightarrow
 L(x,\omega |\varphi ),\qquad
 L(x+\varphi ,\omega|\varphi  )=\frac{1}{2}
 \int_{|x|}^{\infty }dy\,\,{\rm e}\,^{-\omega y}\Gamma (x,y|\varphi ).
\end{equation}
The integral converges for $\mathop{\mathrm{Re}}\omega $ sufficiently large, when the Laplace exponential can beat the growth of $\Gamma $. The shift in $x$  and the factor of $\frac12$ are introduced for later notational convenience. The function $L(x,\omega |\varphi )$ is analytic in $\omega $, at least when $\mathop{\mathrm{Re}}\omega $ is large enough. The inverse transform is
\begin{equation}
 \Gamma (x,y|\varphi )=\int_{C-i\infty }^{C+i\infty }\frac{d\omega }{\pi i}\,\,
 \,{\rm e}\,^{\omega y}L(x+\varphi ,\omega |\varphi ),
\end{equation}
where the contour lies at the right of all the singularities of $L$. The rightmost singularity, which we denote by $\omega _0$, reflects the exponential growth of $\Gamma $ at large $y$. At any fixed $\omega $, $L(x,\omega |\varphi )$ exponentially decreases at $x\rightarrow \pm\infty $ and thus admits a well-defined Fourier transform.

By changing the order of integration, one can show  that  for any function ${\cal R}(s)$,
\begin{eqnarray}
 \int_{0}^{s}ds'\,\mathcal{R}(s')\Gamma (t,s-s'|\varphi )&\rightarrow &
 \hat{\mathcal{R}}\left(\omega +\frac{\partial }{\partial x}\right)L(x,\omega |\varphi )
\nonumber \\
\int_{0}^{t}dt'\,\mathcal{R}(t')\Gamma (t-t',s|\varphi )&\rightarrow &
\hat{\mathcal{R}}\left(\omega -\frac{\partial }{\partial x}\right)L(x,\omega |\varphi ).
\end{eqnarray}
In these formulas $\hat{\mathcal R}$ stands for the Laplace transform of the function $\mathcal R$. We now define the operator $D_t$ such that\footnote{The delta function is defined to give $1$ upon integration from zero, in this sense it corresponds to $\delta (t-0 )$.}:
\begin{equation}
 D_tW(t)=\delta (t),
\end{equation}
so that its Laplace transform is\footnote{Here and in the following we omit the symbol\ \ $\hat{}$\ \ to refer to the Laplace transform of a function in those cases where it is evident from the context.}
\begin{equation}\label{Domega}
 D(\omega )=\frac{1}{W(\omega )}=\frac{\omega +\sqrt{\omega ^2-4g}}{2}.
\end{equation}
At $g=0$, $D_t$ coincides with the ordinary derivative.
Applying $D_tD_s$ to both sides of (\ref{improvedDysonG}), we find:
\begin{equation}\label{Diff-Dyson}
 D_tD_s\Gamma (t,s|\varphi )-G(\varphi +s-t)\Gamma (t,s|\varphi )=\delta (t)\delta (s),
\end{equation}
which, upon the Laplace transform, becomes
\begin{equation}
 \left(D\left(\omega -\frac{\partial }{\partial x}\right)D\left(\omega +\frac{\partial }{\partial x}\right)
 -G(x)
 \right)L(x,\omega |\varphi )=\delta (x-\varphi ).
\end{equation}

This chain of arguments shows that $L(x,\omega |\varphi )$ is the Green's function of a particle with the dispersion relation $\varepsilon (p)=D(\omega +ip)D(\omega -ip)$ moving in a $2\pi$-periodic potential $-G(x)$. Such a quantum-mechanical problem has a band spectrum, the eigenfunctions have Bloch form $\,{\rm e}\,^{ipx}\psi _n(x)$ with $2\pi $-periodic $\psi _n$ and quasimomentum $p$ constrained to the Brillouin zone $-1/2<p<1/2$. The eigenfunctions are solutions of the Schr\"odinger equation
\begin{equation}\label{schreq}
 \left(D\left(\omega -ip-\frac{\partial }{\partial x}\right)D\left(\omega +ip+\frac{\partial }{\partial x}\right)-G(x)\right)\psi _n(x,\omega ;p)=E_n(\omega ;p)\psi _n(x,\omega ;p).
\end{equation}
In consequence, $L(x,\omega |\varphi )$ admits the following spectral representation in terms of the solutions to the Schr\"odinger equation:
\begin{equation}
 L(x,\omega |\varphi )=\sum_{n}^{}\int_{-\frac{1}{2}}^{\frac{1}{2}}dp\,
 \,{\rm e}\,^{ip(x-\varphi )}\,
 \frac{\psi_n ^*(\varphi ,\omega ;p)\psi _n(x,\omega ;p)}{E_n(\omega ;p)}\,.
\end{equation}
From (\ref{kW1W2}) we then get the  spectral representation of the Wilson loop correlator:
\begin{eqnarray}\label{spectral}
 &&\left\langle W(C_1)W(\bar{C}_2)\right\rangle_{\rm ladders} =
 \int_{C-i\infty }^{C+i\infty }\frac{d\omega }{\pi i}\,\,\,{\rm e}\,^{4\pi \omega }
 \sum_{n}^{}\int_{-\frac{1}{2}}^{\frac{1}{2}}dp\,\,\frac{1}{E_n(\omega ;p)}
\nonumber \\
 &&\times
 \int_{0}^{2\pi }d\varphi \,
 \psi _n^*(\varphi ,\omega ;p)V\left(\omega +ip+\frac{\partial }{\partial \varphi }\right)G(\varphi )\psi _n(\varphi ,\omega ;p).
\end{eqnarray}
This differs from the result in \cite{Zarembo:2001jp} by an insertion of the operator $V$. As explained above (see also \cite{Correa:2018lyl}),  this insertion takes into account different combinatorics of the ladder diagrams in the two loops correlator compared to a single Wilson loop.

\subsubsection{Strong coupling limit}

When the coupling is large, $g$ and $G_{ab}$ go to infinity simultaneously. The spectral representation for the Wilson loop correlator then features strong exponential enhancement. Indeed, the $\omega $ integral in (\ref{spectral}) is saturated by the rightmost singularity of the integrand. The exponential behavior of the Wilson loop correlator is governed by the position of this singularity:
\begin{equation}
\left\langle W(C_1)W(\bar{C}_2)\right\rangle_{\rm ladders} \simeq \,{\rm e}\,^{4\pi \omega _0},
\end{equation}
as long as $\omega _{0}$ goes to infinity at strong coupling.

There are actually two possible scenarios. Both $V(\omega)$ and $D(\omega)$ have a square-root branch point at
\begin{equation}\label{worainbows}
\omega^{r} _0=2\sqrt{g}\,.
\end{equation}
This singularity appears in the expectation value of a single Wilson loop. As such, it reflects combinatorics of rainbow diagrams. The
branch point at $\omega =\omega^{r} _{0}$ affects the integrand in the spectral representation through the kernel $V(\omega )$ and also through the eigenfunctions and eigenvalues of the Schr\"odinger equation (\ref{schreq}), which inherit this singularity from the function $D(\omega )$ in the kinetic energy.

If no other singularities lie to the right of $\omega _{0}^{r}$, the branch point at $\omega =\omega^{r} _{0}$ dictates the strong-coupling asymptotics of the correlator. In that case,
\begin{equation}\label{connected-strong}
\left\langle W(C_1)W(\bar{C}_2)\right\rangle_{\rm ladders} \simeq \,{\rm e}\,^{2\sqrt{\lambda }}\simeq \left\langle W(C)\right\rangle^{2}.
\end{equation}
The main contribution to the correlator then comes from disconnected diagrams without exchanges between the two loops. The exchange, ladder diagrams are statistically less numerous than rainbow diagrams, and the connected correlator behaves as the square of the Wilson loop expectation value.

Other possible singularities of the integrand in (\ref{spectral}) are cuts associated with the Brillouin zones. At the bottom of a Brillouin zone, the energy is quadratic in quasi-momentum:
\begin{equation}
 E_{n}(\omega ,p)=E_{n}(\omega ,0)+\frac{1}{2}\,E''_{n}(\omega ,0)p^{2}+\ldots
\end{equation}
The momentum integration produces a branch cut when the zone boundary cros\-ses zero.
The rightmost singularity corresponds to the bottom of the lowest zone:
\begin{equation}\label{zoneE=0}
E_0(\omega^{l} _0;0)=0.
\end{equation}

In the strong-coupling limit  the Schr\"o\-din\-ger problem  (\ref{schreq}) becomes semi-classical (see \cite{Zarembo:2001jp} for a detailed justification), and the bottom of the lowest zone coincides with the minimum of the classical energy,
given by
\begin{equation}\label{class-E}
E_0(\omega ;0)\simeq D^2(\omega )-G(0).
\end{equation}
The condition for the zero crossing is
\begin{equation}
 D(\omega_{0}^{l} )=\sqrt{G(0)}.
\end{equation}
The function $D(\omega )$ is given by (\ref{Domega}) and takes positive real values on the semi-infinite interval $\omega >2\sqrt{g}$, growing monotonously from $D(2\sqrt{g})=\sqrt{g}$ to infinity. Hence, there are two possible scenarios: (i) ${G(0)}<g$, the equation for $\omega _{0}^{l}$ then has no solutions, and (ii) ${G(0)}>g$, then
\begin{equation}\label{woladders}
\omega ^{l}_0=\sqrt{G(0)}+\frac{g}{\sqrt{G(0)}}\,,\qquad \left({G(0)}>g\right),
\end{equation}
 such that $\omega _{0}^{l}$ is always larger than $\omega _{0}^{r}=2\sqrt{g}$.

Competition between the two singular points (\ref{worainbows}) and (\ref{woladders}) determines the phase structure of the correlator.  If the solution (\ref{woladders}) exists, $\omega _{0}^{l}$ always constitutes the leading singularity. The correlator is then saturated by the ladder diagrams. The singular point $\omega _{0}^{l}$ collides with $\omega _{0}^{r}$ and moves under the cut once ${G(0)}$ reaches $g$. Beyond that point, the rainbow graphs are more important than ladder exchanges between the two loops. The two regimes are separated by a phase transition, which is analogous to the Gross-Ooguri transition between connected and disconnected minimal surfaces  in string theory.

\begin{figure}[t]
\begin{center}
 \centerline{\includegraphics[width=7cm]{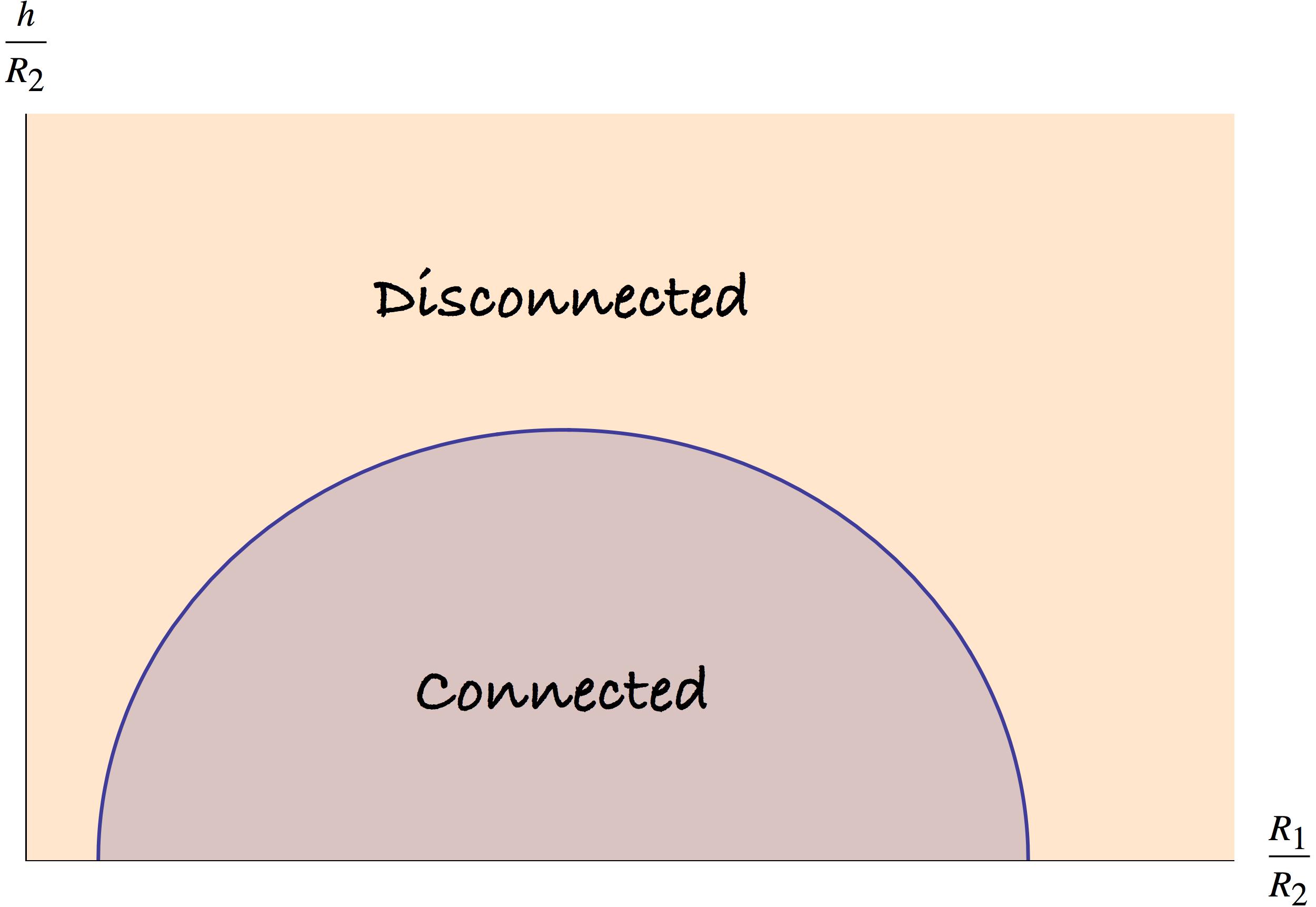}}
\caption{\label{WbarW}\small The phase diagram for two loops of opposite orientation.}
\end{center}
\end{figure}

The transition happens when $G(0)=g$.
Taking into account the explicit form of the ladder propagator (\ref{G(theta)}) we find the critical separation between the two loops:
\begin{equation}
h_c=\sqrt{2R_1R_2\left(1+\cos\gamma \right)-\left(R_1-R_2\right)^2}\,.
\label{hcritical}
\end{equation}
The resulting phase diagram for  $\cos\gamma =1$ is shown in fig.~\ref{WbarW}. When $R_1=R_2\equiv R$, we get $h_c=2R$, in agreement with \cite{Zarembo:2001jp}. As $\cos\gamma \rightarrow -1$, the connected region shrinks to a point -- in this extreme case rainbow diagrams always give the dominant contribution to the Wilson loop correlator.

\begin{figure}[t]
\begin{center}
 \subfigure[]{
 \begin{tikzpicture}[scale=0.73]
\node at (0,0) {\includegraphics[width=5cm]{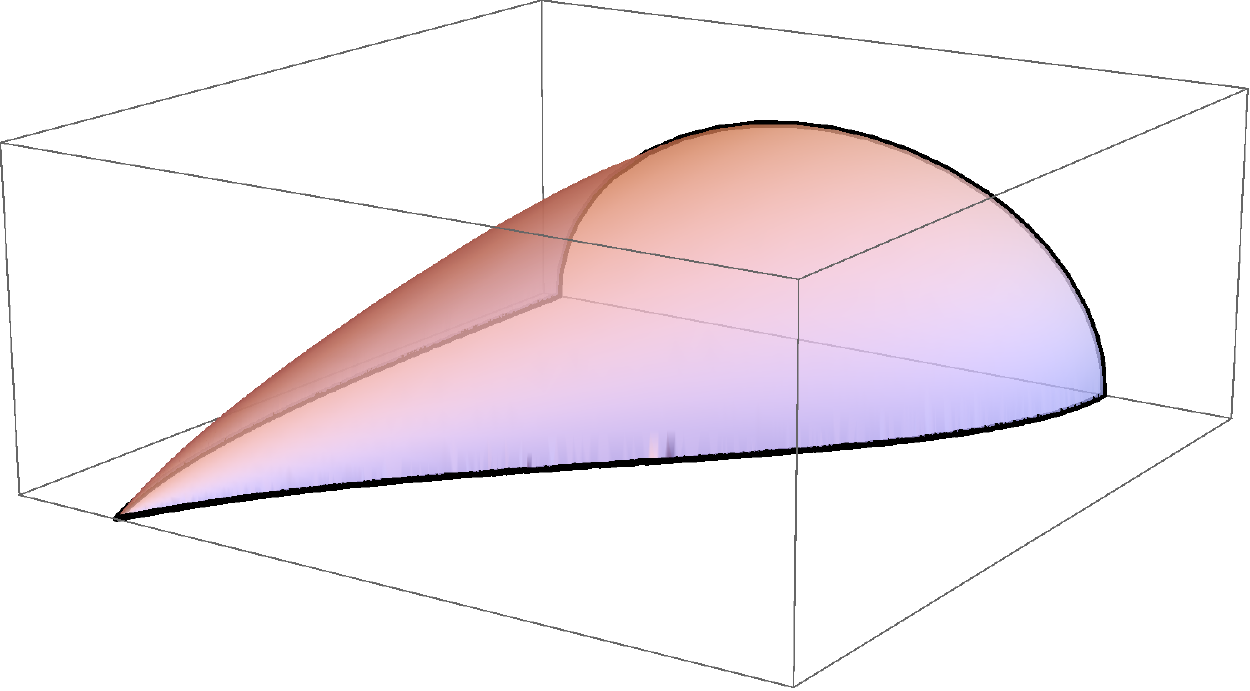}};
\draw (-2.81,-1.45) node {\scriptsize $1$};
\draw (-3.45,-1.25) node {\scriptsize $0$};
\draw (0.9,-2.45) node {\scriptsize $7$};
\draw (-1.7,-2.5) node {\footnotesize $\frac{R_1}{R_2}$};
\draw (-2.81,-0.95)--(-2.81,-0.8);
\draw (-3.85,-0.8) node {\scriptsize $0$};
\draw (-3.85,1.1) node {\scriptsize $4$};
\draw (-4.6,0.1) node {\footnotesize $\frac{h}{R_1}$};
\draw (-3.6,1.45) node {\scriptsize $\pi$};
\draw (-0.5,2.3) node {\scriptsize $0$};
\draw (-2.5,2.4) node {\footnotesize $\gamma$};
  \end{tikzpicture}
   \label{3DPhD:subfig1}
 }
 \hspace{0.1cm}
 \subfigure[]{
  \begin{tikzpicture}[scale=0.73]
\node at (0,0) {\includegraphics[width=5cm]{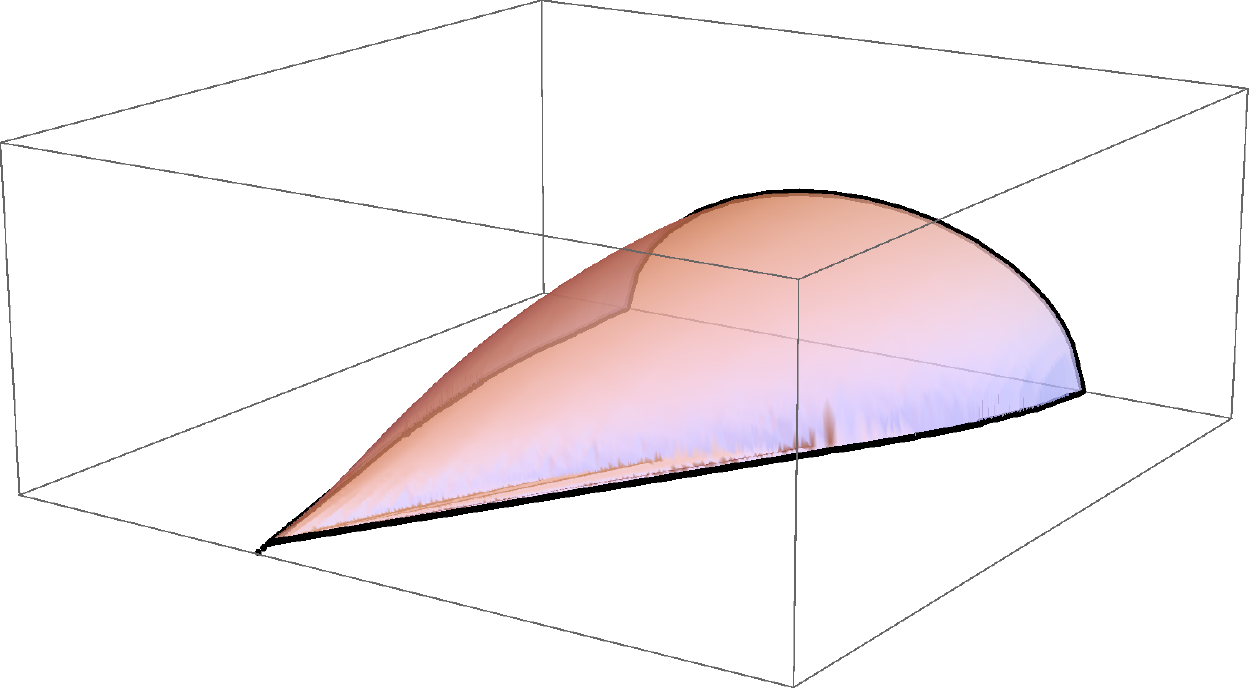}};
\draw (-2.3,-1.68) node {\scriptsize $1$};
\draw (-3.45,-1.25) node {\scriptsize $0$};
\draw (0.9,-2.45) node {\scriptsize $4$};
\draw (-1.7,-2.5) node {\footnotesize $\frac{R_1}{R_2}$};
\draw (-2.03,-1.15)--(-2.03,-1);
\draw (-3.85,-0.8) node {\scriptsize $0$};
\draw (-3.85,1.1) node {\scriptsize $2$};
\draw (-4.6,0.1) node {\footnotesize $\frac{h}{R_1}$};
\draw (-3.6,1.45) node {\scriptsize $\pi$};
\draw (-0.5,2.3) node {\scriptsize $0$};
\draw (-2.5,2.4) node {\footnotesize $\gamma$};
    \end{tikzpicture}
   \label{3DPhD:subfig2}
 }
 \caption{\small Comparison between ladders contribution phase transition and the Gross-Ooguri phase transition}
 \end{center}
\end{figure}
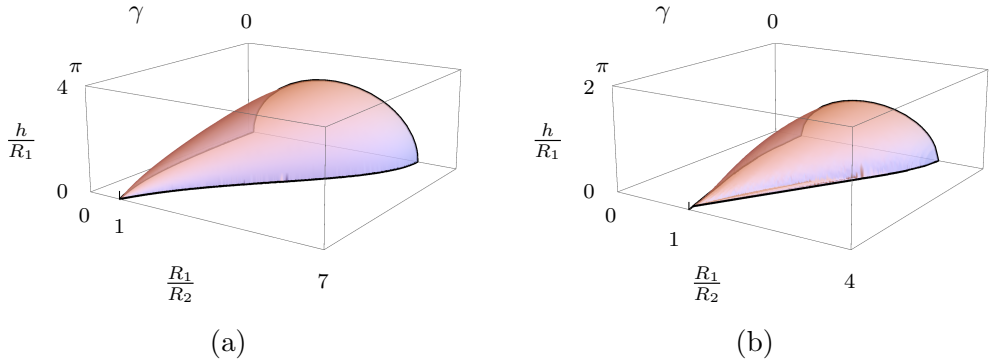

The transition happens even if $h=0$. The connected phase then exists for
\begin{equation}
2+\cos\gamma -\sqrt{\left(1+\cos\gamma \right)\left(3+\cos\gamma \right)}<\frac{R_1}{R_2}<2+\cos\gamma +\sqrt{\left(1+\cos\gamma \right)\left(3+\cos\gamma \right)}\,.
\end{equation}
In fact, eq. (\ref{hcritical}) specifies a region in a 3-dimensional diagram with axes $\frac{h}{R_2}$, $\frac{R_1}{R_2}$ and $\gamma$. The interior of the purple surface in fig. \ref{3DPhD:subfig1} corresponds to the region of parameters where the ladder diagrams dominate over rainbow diagrams. Remarkably, and despite the contribution of interaction diagrams to the correlator has been omitted, ladder diagrams capture all the qualitative features of the Gross-Ooguri phase transition. The latter is represented in the fig. \ref{3DPhD:subfig2}, using the solution found in \cite{Correa:2018lyl}. The region under the purple surface in this plot represents the configurations in which the area of the connected dual worldsheet is the minimal one.

This is consistent with the picture of the correlator saturated by the dense net of ladder or rainbow diagrams, depending on the spacial arrangement of the two contours.

It is perhaps worthwhile to give an alternative, simplified derivation of the strong-coupling behavior that lacks rigor, but instead is more physically transparent. The Dyson equation (\ref{Diff-Dyson}) can be formally written as
\begin{equation}\label{ddieq}
 \left(D\left(\frac{\partial }{\partial y}+\frac{\partial }{\partial x}\right)
 D\left(\frac{\partial }{\partial y}-\frac{\partial }{\partial x}\right)-G(\varphi +x)
 \right)\Gamma (x,y|\varphi )=2\delta (x)\delta (y),
\end{equation}
where $D(\omega )$ is given by (\ref{Domega}).
Anticipating an exponential growth of $\Gamma $ we  look for a solution of the form
\begin{equation}\label{expGamma}
 \Gamma (x,y|\varphi )\sim \psi (x)\,{\rm e}\,^{\Omega y}.
\end{equation}
Substituting this ansatz into (\ref{ddieq}) we find:
\begin{equation}
 \left(D\left(\Omega +\frac{\partial }{\partial x}\right)
 D\left(\Omega -\frac{\partial }{\partial x}\right)-G(\varphi +x)
 \right)\psi (x)=0.
\end{equation}
This can be viewed as an eigenvalue equation for $\Omega $, which is essentially equivalent to (\ref{schreq}) with zero energy and quasi-momentum. At strong coupling $G$, $\Omega ^{2}$ and $D^{2}$ all scale as $g\sim \lambda $. The problem becomes semiclassical, and the maximal possible eigenvalue $\Omega $
is determined by a classical computation where we look for a solution of
\begin{equation}
 D^{2}(\Omega )-G(0)=0,
\end{equation}
taking into account that  $G(\varphi +x )$ reaches maximum at zero. The solution to this equation exists only for $G(0)>g$ and then  is given in (\ref{woladders}). For $G(0)<g$ we have to take $\Omega =2\sqrt{g}$, the smallest value allowed by analyticity of the kinetic energy.

Upon substituting (\ref{expGamma}) into (\ref{kW1W2}), we get, keeping  an exponential accuracy:
\begin{equation}
 \left\langle W(C_1)W(\bar{C}_2)\right\rangle_{\rm ladders}=K(2\pi )\sim
 \int_{0}^{2\pi }dt'\,\,{\rm e}\,^{2\sqrt{g}\left(2\pi -t'\right)+\Omega (2\pi +t')}.
\end{equation}
If $\Omega >2\sqrt{g}$, the main contribution to the integral comes from $t'\sim 2\pi $ and is determined by the asymptotics of $\Gamma (x,y|\varphi )$. While for $\Omega =2\sqrt{g}$, all the interval of integration contributes, and we get the asymptotic behavior (\ref{connected-strong}) dictated by disconnected diagrams. The transition between the two regimes happens when $G(0)=g$.

\subsection{Strong coupling limit for same orientation}

For loops of the same orientation the change of variables from $s$ and $t$ to $x$ and $y$ results in
\begin{equation}\label{tilde-ddieq}
 \left(D\left(\frac{\partial }{\partial y}+\frac{\partial }{\partial x}\right)
 D\left(\frac{\partial }{\partial y}-\frac{\partial }{\partial x}\right)-\widetilde{G}(\varphi -y)
 \right)\widetilde{\Gamma }(x,y|\varphi )=2\delta (x)\delta (y).
\end{equation}
The potential now depends on $y$ and to the first approximation we can just neglect  the $x$ dependence. A natural ansatz to start with is
\begin{equation}
 \widetilde{\Gamma }(x,y|\varphi )\sim \,{\rm e}\,^{S(y)}.
\end{equation}
Denoting
\begin{equation}
 \Omega (y)=S'(y),
\end{equation}
we get in the semiclassical limit:
\begin{equation}
 D^{2}(\Omega )-\widetilde{G}(\varphi -y)=0.
\end{equation}
Again, this is solved by
\begin{equation}\label{om}
 \Omega (y)=\sqrt{\widetilde{G}(\varphi -y)}+\frac{g}{\sqrt{\widetilde{G}(\varphi -y)}}\,,
\end{equation}
for $\widetilde{G}>g$ and we should take $\Omega =2\sqrt{g}$ for $\widetilde{G}<g$. In either case, the action $S$ scales as $\sqrt{\lambda }$ which justifies the use of the semiclassical approximation at strong coupling.
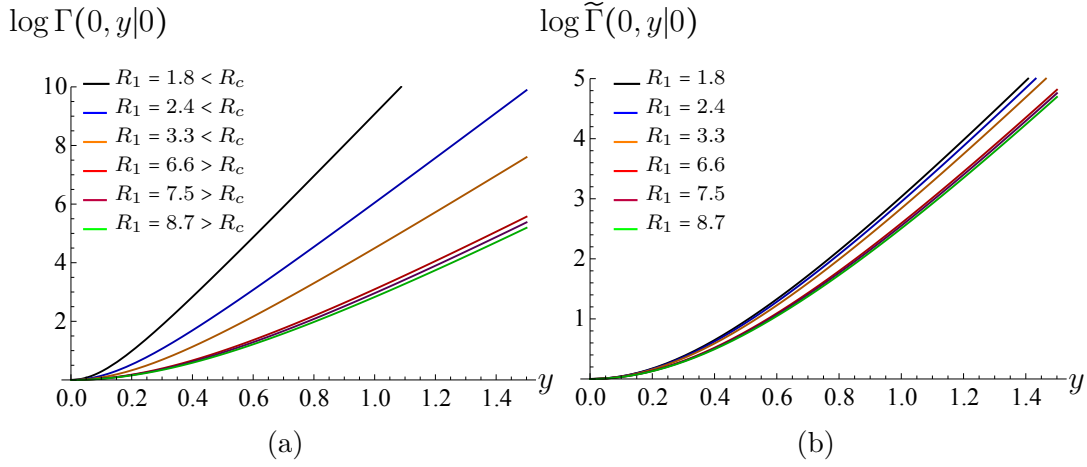
\begin{figure}
			\centering     
			\subfigure[]{\label{logG-y:subfig1}\includegraphics[width=0.48\linewidth]{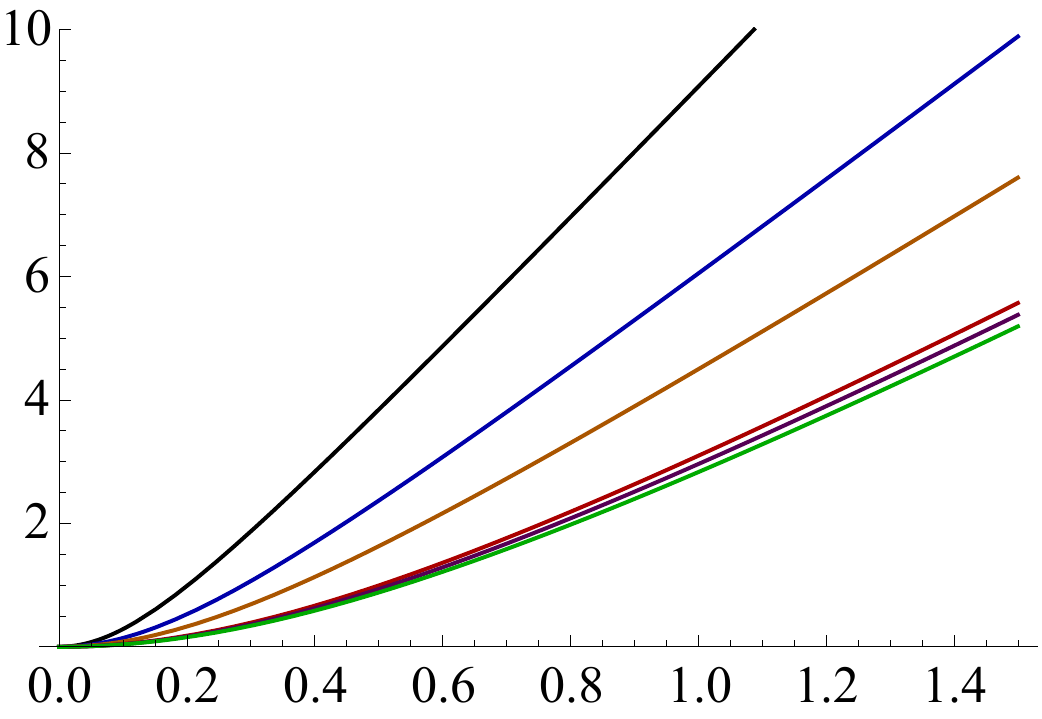}
            \put(-200,145){$\log\Gamma (0, y\vert 0) $}
            \put(0,10){$y$}
            \put(-160,125){\scriptsize $R_1=1.8 < R_c$}
            \put(-160,114){\scriptsize $R_1=2.4 < R_c$}
            \put(-160,103){\scriptsize $R_1=3.3 < R_c$}
            \put(-160,92){\scriptsize $R_1=6.6 > R_c$}
            \put(-160,81){\scriptsize $R_1=7.5 > R_c$}
            \put(-160,70){\scriptsize $R_1=8.7 > R_c$}
            \thicklines
            \put(-172,125){\line(1,0){10}}
            \put(-172,114){\color{blue}\line(1,0){9}}
            \put(-172,103){\color{orange}\line(1,0){9}}
            \put(-172,92){\color{red}\line(1,0){9}}
            \put(-172,81){\color{purple}\line(1,0){9}}
            \put(-172,70){\color{green}\line(1,0){9}}}
			\hspace{0.2cm}
			\subfigure[]{\label{logG-y:subfig2}\includegraphics[width=0.48\linewidth]{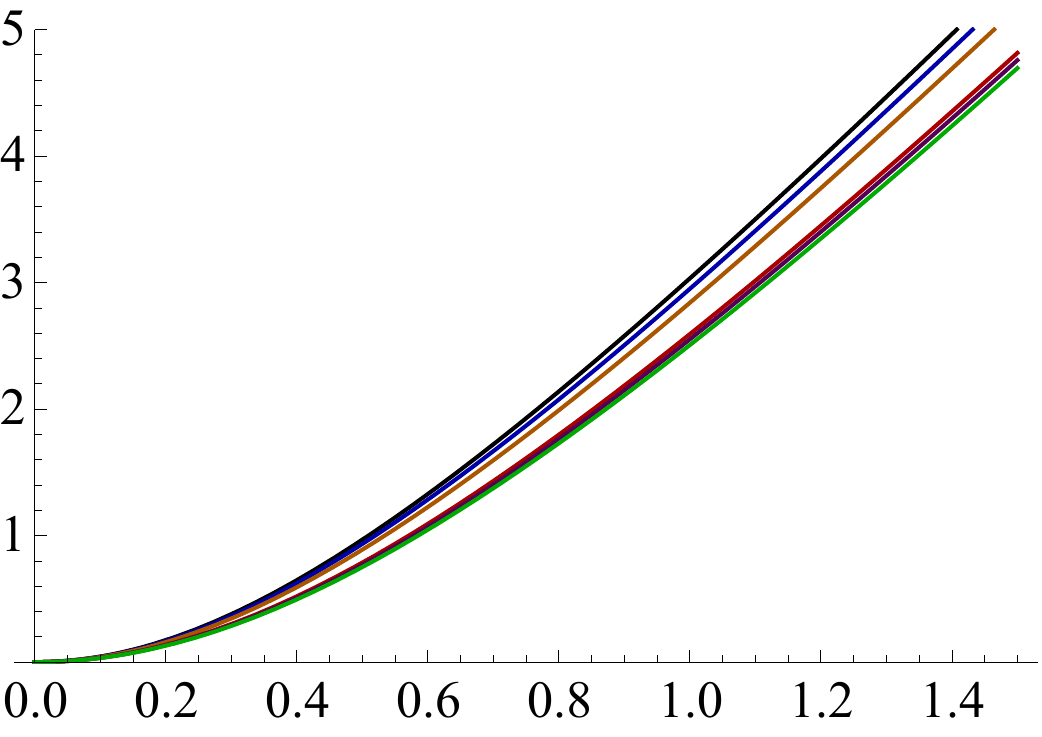}
            \put(-200,145){$\log\widetilde\Gamma (0, y\vert 0) $}
            \put(0,10){$y$}
            \put(-160,125){\scriptsize $R_1=1.8$}
            \put(-160,114){\scriptsize $R_1=2.4$}
            \put(-160,103){\scriptsize $R_1=3.3$}
            \put(-160,92){\scriptsize $R_1=6.6 $}
            \put(-160,81){\scriptsize $R_1=7.5 $}
            \put(-160,70){\scriptsize $R_1=8.7 $}
            \thicklines
            \put(-172,125){\line(1,0){10}}
            \put(-172,114){\color{blue}\line(1,0){9}}
            \put(-172,103){\color{orange}\line(1,0){9}}
            \put(-172,92){\color{red}\line(1,0){9}}
            \put(-172,81){\color{purple}\line(1,0){9}}
            \put(-172,70){\color{green}\line(1,0){9}}}
\caption{\label{logG-y}\small The $y$ dependence of the Bethe-Salpeter wavefunction from numerical solution of the Dyson equation for various values of $R_{1}$. The other parameters are set to $h=0$, $R_{2}=1$ and $g=10$: (a) for loops of opposite orientation and $\gamma =0$. For these values of parameters the Gross-Ooguri transition happens at $R_{c}=3+2\sqrt{2}\simeq 5.83$; (b) for loops of the same orientation and $\gamma =\pi /4$.}
\end{figure}

The strong-coupling estimate of the Wilson loop correlator is
\begin{equation}\label{veryapprox}
 \left\langle W(C_1)W({C}_2)\right\rangle_{\rm ladders}\sim
 \int_{0}^{2\pi }dt'\,\,{\rm e}\,^{2\sqrt{g}\left(2\pi -t'\right)+S (2\pi +t')}.
\end{equation}
The ladder diagrams would give the dominant contribution if the integral were saturated by a non-trivial saddle-point:
\begin{equation}
 S'(2\pi +t_{*})=2\sqrt{g}\,.
\end{equation}
Since $S'=\Omega $, and $\Omega $ is given by (\ref{om}) the saddle-point condition becomes
\begin{equation}\label{spcond-same}
 \widetilde{G}(\theta _{*})=g.
\end{equation}
However this scenario is never realized for real values of the parameters, because
\begin{equation}\label{inequality}
 \widetilde{G}(\theta )\leq \widetilde{G}(\pi )= 2gR_{1}R_{2}\,\frac{1+\cos\gamma }{(R_{1}+R_{2})^{2}+h^{2}}< g,
\end{equation}
and the saddle-point condition (\ref{spcond-same}) never has a solution.

\begin{figure}
			\centering     
			 \subfigure[]{\label{Om:subfig1}\includegraphics[width=0.48\linewidth]{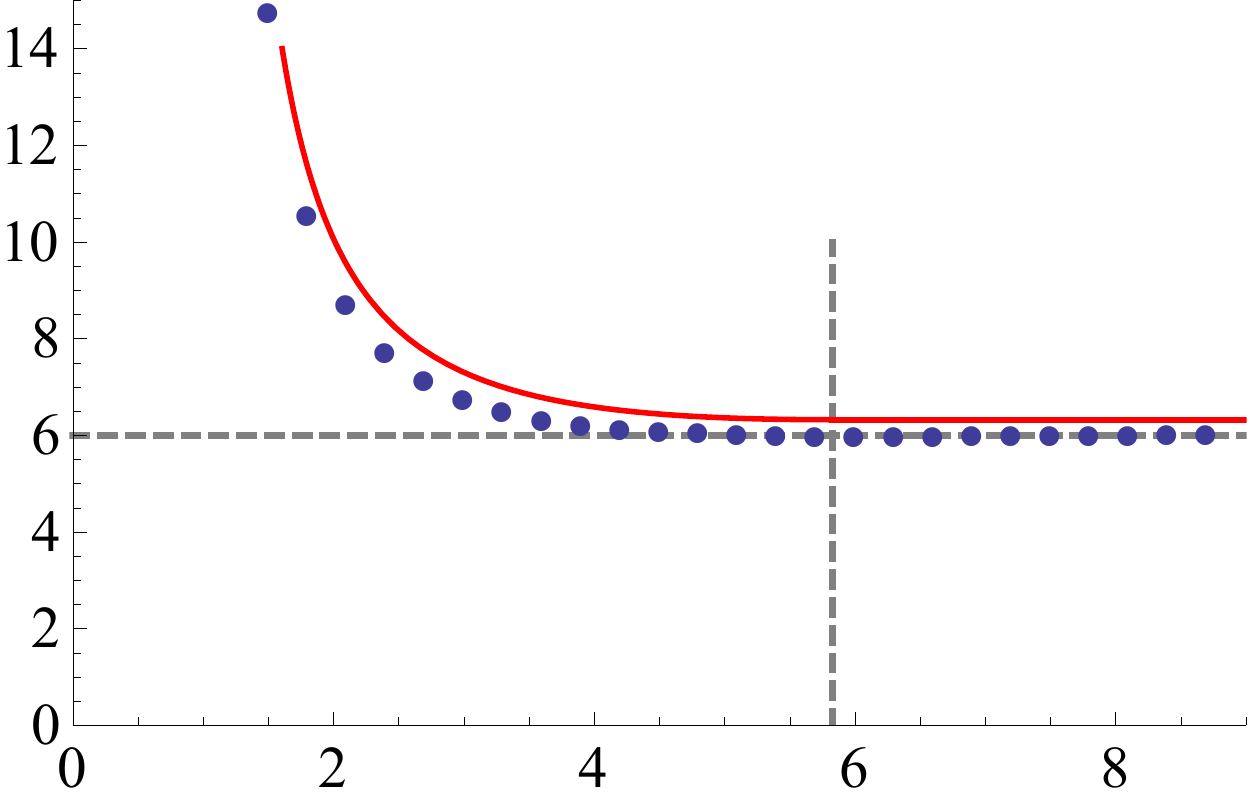}\put(-175,125){$\Omega$}\put(-2.5,0){$R_1$}\put(-80,18){$R_c$}}
			\hspace{0.1cm}
			\subfigure[]{\label{Om:subfig2}\includegraphics[width=0.48\linewidth]{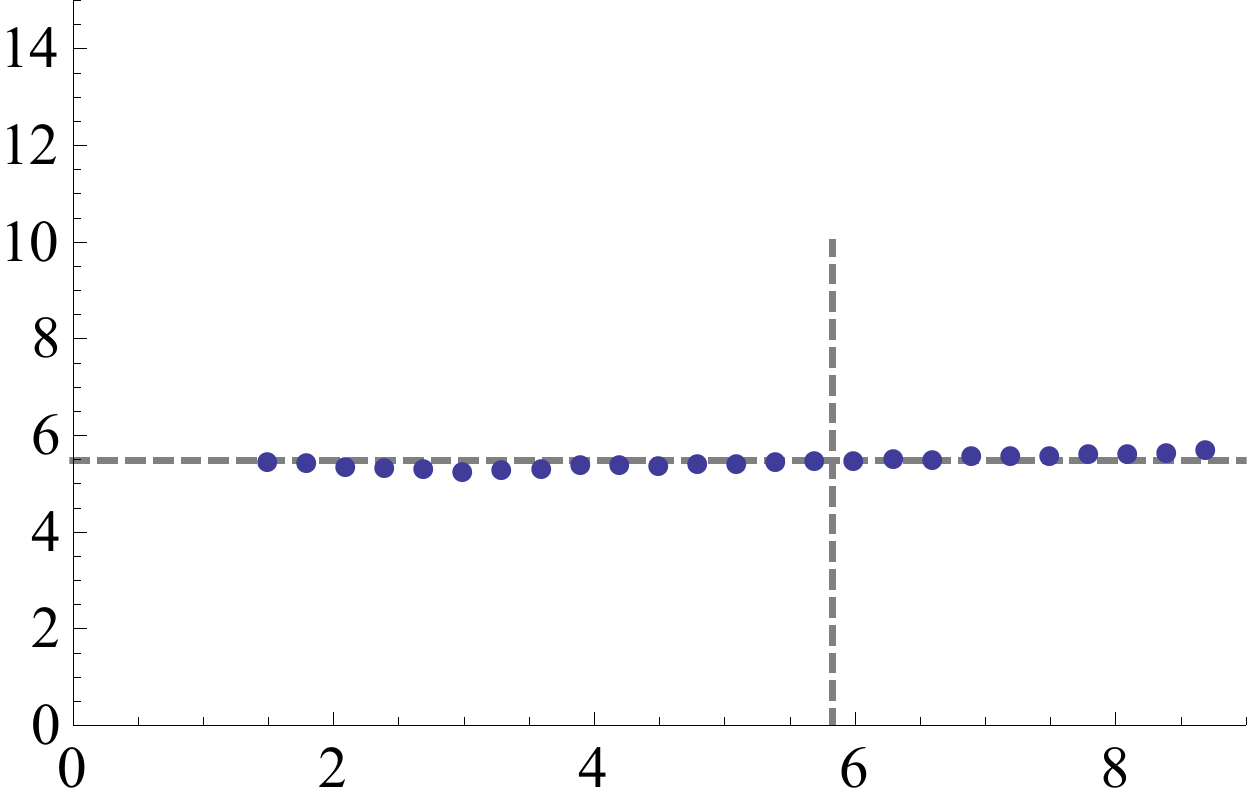}\put(-175,125){$\Omega$}\put(-2.5,0){$R_1$}}
\caption{\label{Om}\small The exponent in (\ref{expGamma}) extracted from numerical. The parameters take the same values as in fig.~\ref{logG-y}: (a) The Gross-Ooguri transition is clearly visible for opposite-orientation loops. It is clear from the plot that the transition is second order. The red curve corresponds to the analytical result for (\ref{worainbows}) and (\ref{woladders}) and the difference with the numerical is attributed to finite $g$ effects ; (b) There is no phase transition for loops of the same orientation.}
\end{figure}
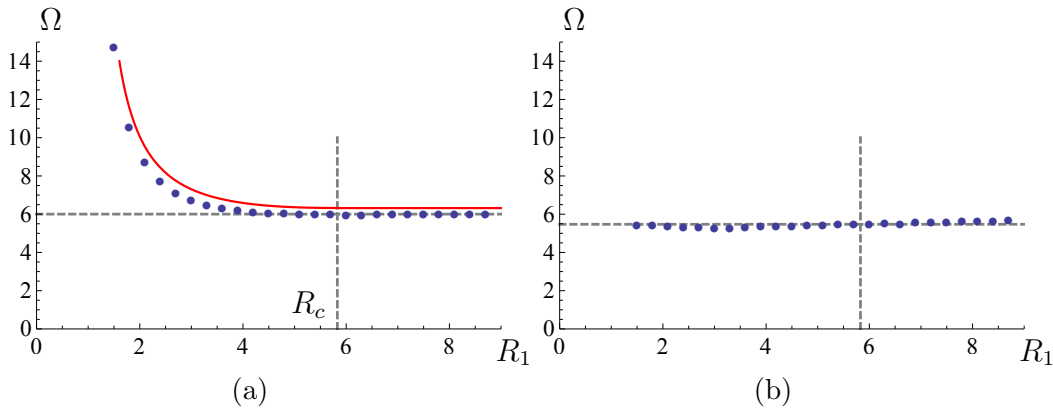

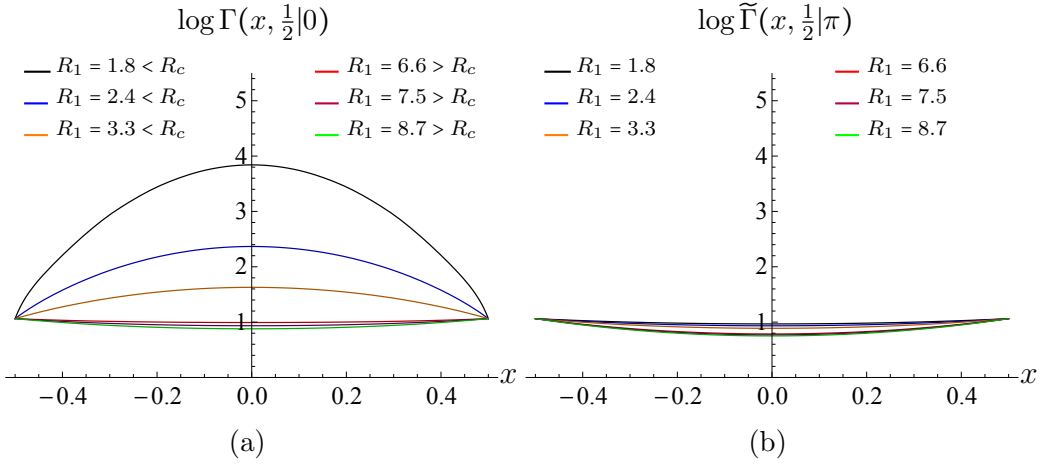
\begin{figure}[t]
			\centering     
			\subfigure[]{\label{logGx:subfig1}\includegraphics[width=0.48\linewidth]{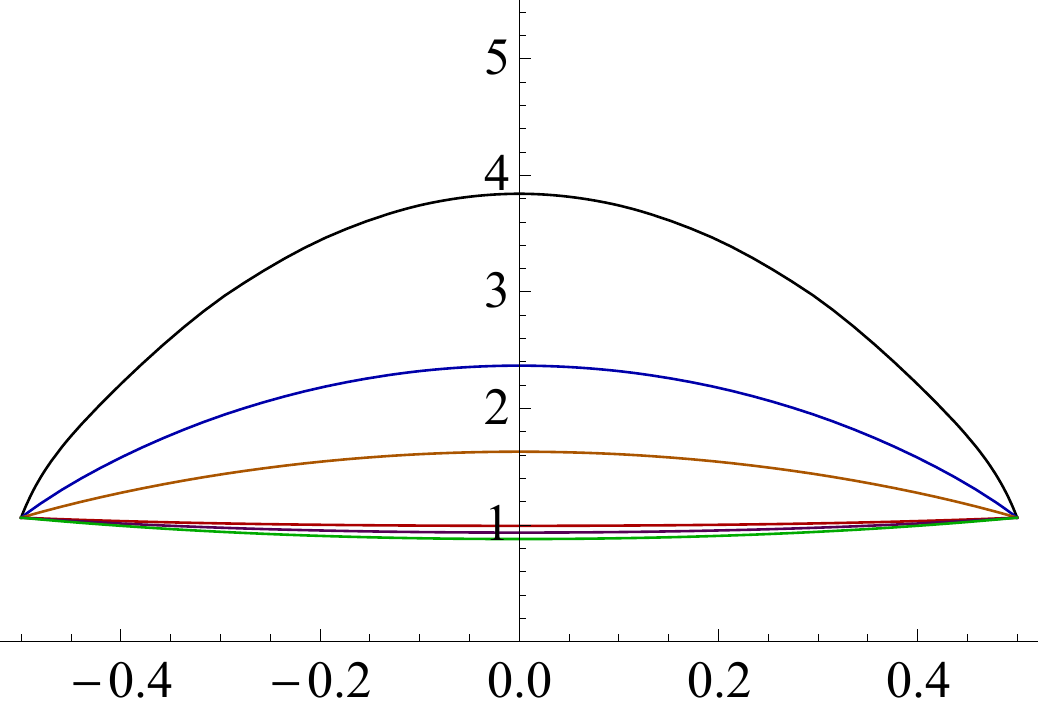}
            \put(-122,145){\small $\log\Gamma (x, \tfrac{1}{2} \vert 0) $}
            \put(0,11){$x$}}
            \put(-168,129){\scriptsize $R_1=1.8 < R_c$}
            \put(-168,117){\scriptsize $R_1=2.4 < R_c$}
            \put(-168,105){\scriptsize $R_1=3.3 < R_c$}
            \put(-58,129){\scriptsize $R_1=6.6 > R_c$}
            \put(-58,117){\scriptsize $R_1=7.5 > R_c$}
            \put(-58,105){\scriptsize $R_1=8.7 > R_c$}
            \thicklines
            \put(-180,129){\line(1,0){10}}
            \put(-180,117){\color{blue}\line(1,0){9}}
            \put(-180,105){\color{orange}\line(1,0){9}}
            \put(-70,129){\color{red}\line(1,0){9}}
            \put(-70,117){\color{purple}\line(1,0){9}}
            \put(-70,105){\color{green}\line(1,0){9}}
			\hspace{0.2cm}
			\subfigure[]{\label{logGx:subfig2}\includegraphics[width=0.48\linewidth]{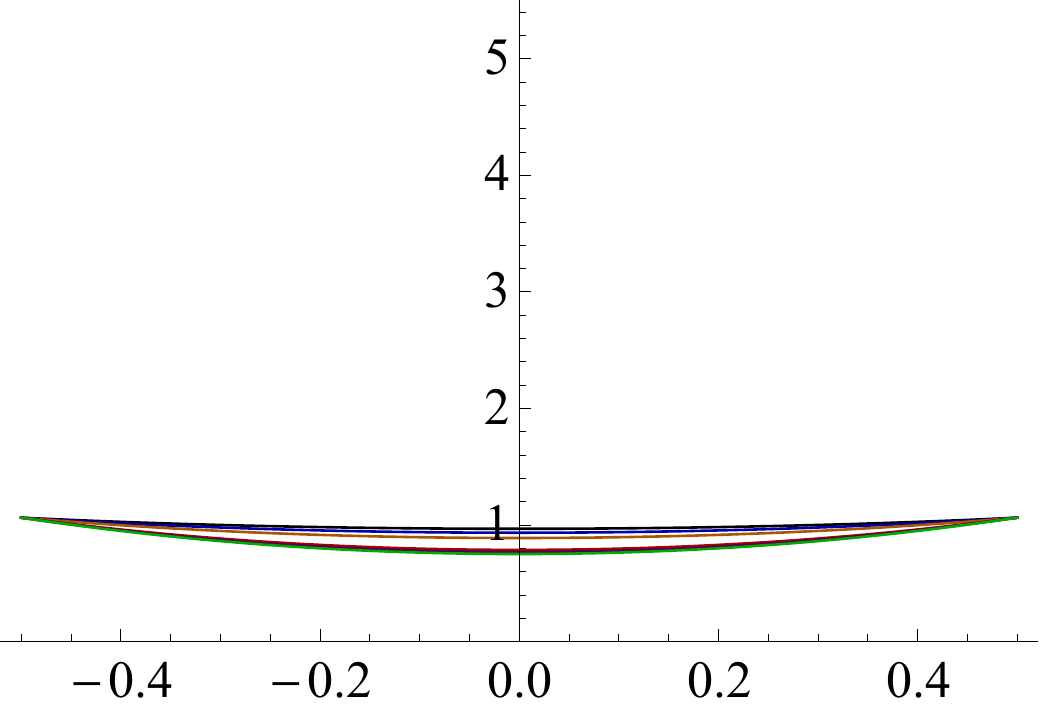}
 \put(-122,145){\small $\log\widetilde\Gamma (x, \tfrac{1}{2} \vert \pi) $}
            \put(0,11){$x$}}
            \put(-168,129){\scriptsize $R_1=1.8 $}
            \put(-168,117){\scriptsize $R_1=2.4 $}
            \put(-168,105){\scriptsize $R_1=3.3 $}
            \put(-58,129){\scriptsize $R_1=6.6 $}
            \put(-58,117){\scriptsize $R_1=7.5 $}
            \put(-58,105){\scriptsize $R_1=8.7 $}
            \thicklines
            \put(-180,129){\line(1,0){10}}
            \put(-180,117){\color{blue}\line(1,0){9}}
            \put(-180,105){\color{orange}\line(1,0){9}}
            \put(-70,129){\color{red}\line(1,0){9}}
            \put(-70,117){\color{purple}\line(1,0){9}}
            \put(-70,105){\color{green}\line(1,0){9}}
\caption{\label{logGx}\small Dependence of the Bethe-Salpeter wavefunction on $x$ at fixed $y$. The parameters are the same as in fig.~\ref{logG-y}: (a) in the ladder phase $\psi (x)$ in (\ref{expGamma}) has a clearly pronounced profile, while in the rainbow phase the dependence on $x$ is almost flat (b) The dependence on $x$ is much weaker for loops of the same orientation.}
\end{figure}

We thus conclude that the same-orientation correlator is always saturated by the rainbow-type diagrams, and does not undergo the Gross-Ooguri transition. We have checked this picture numerically. The Bethe-Salpeter wavefunction indeed grows exponentially with $y$ at fixed $x$, in agreement with (\ref{expGamma}), as clear from fig.~\ref{logG-y}.  In the ladder phase, the rate of growth $\Omega $ varies with the parameters of the problem (in the numerics we varied $R_{1}$ with all other parameters fixed), as shown in fig.~\ref{Om}. For contours of the same orientation $\Omega $ remains approximately constant. Perhaps the most dramatic manifestation of the phase transition is the change in the $x$ dependence of the Bethe-Salpeter wavefunction, fig.~\ref{logGx}. The dependence on $x$ becomes almost flat in the rainbow phase. The residual, slow variation with $x$ can be attributed to the next order in the semiclassical expansion in $1/\sqrt{g}$.

The absence of the phase transition for same-orientation circular loops is consistent with the expectations from AdS/CFT.
One could try to find a connected worldsheet for coincident orientations as a surface of revolution connecting opposite points on the two circles. But such a surface would contain a self crossing point that leads to a conical singularity. Conical singularities are inconsistent with the string equations of motion and are forbidden on minimal surfaces, so the solution with the cylinder topology for this configuration of Wilson loops does not exist for any choice of parameters.
Solutions which connect coaxial circles of the same orientation can be found \cite{Drukker:2005cu}\footnote{It is unclear to us if these solutions are linearly stable.} for Wilson loops non-trivially extended along $S^{5}$, such that the dual string wraps an $S^2 \subset S^5$ thus avoiding self-crossing in $AdS_{5}$.

The transition for the same orientation occurs upon analytic continuation to imaginary $\gamma $:
\begin{equation}
 \gamma =i\alpha .
\end{equation}
The critical imaginary angle is
\begin{equation}
 \cosh\alpha _{c}=\frac{R_{1}^{2}+R_{2}^{2}+h^{2}}{2R_{1}R_{2}}\,.
\end{equation}
For $\alpha <\alpha _{c}$ the maximum of the propagator still occurs at $\theta=\pi $ and the inequality (\ref{inequality}) still holds. But when $\alpha $ exceeds $\alpha _{c}$ the maximum occurs at zero:
\begin{equation}
 \widetilde{G}(0)=g\,\frac{\cosh\alpha -1}{\frac{R_{1}^{2}+R_{2}^{2}+h^{2}}{2R_{1}R_{2}}-1}>g,
\end{equation}
and moreover $\widetilde{G}(\theta )>g$ for any $\theta $, which means that the integral (\ref{veryapprox}) is saturated on the upper limit. The correlator is governed by the ladder contribution with the exponent $S(4\pi )$. This conclusion is consistent with the fact that at large imaginary $\gamma $ the correlator of Wilson loops is saturated by scalar ladder exchanges, which are enhanced by a factor of $\cosh\alpha $ compared to gluon and scalar rainbow diagrams which do not contain exponential factors.

\subsection{Solution for BPS configurations}

As shown in \cite{Correa:2018lyl}, for some specific relation between the geometric parameters and the internal space separation, the correlator of Wilson loops with opposite orientations is supersymmetric. In such cases the propagator (\ref{G(theta)}) becomes constant and an explicit resummation of ladder diagrams, that matched both matrix model computations and the holographic description, is possible.

In this section we first show how this result can be recovered by solving the Dyson equation \eqref{improvedDysonG}, and then extend a similar analysis for the correlator of Wilson loops with equal orientations, {\it i.e.} by solving \eqref{improvedDysonGtilde}
for some other specific critical relation between the parameters.

From expression \eqref{G(theta)}, it is immediate that the critical relation in the case of opposite orientations is
\begin{equation}
\cos\gamma = -\frac{R_1^2+R_2^2+h^2}{2R_1R_2} \quad  \Rightarrow \quad
G(\theta) = -g
\end{equation}

The effective propagator being constant, the integral \eqref{improvedDysonG} becomes a convolution
in both variables $t$ and $s$ with the function $W$. Since $\Gamma$ is independent of $\varphi$, we will omit $\varphi $ to simplify the notations. Thus, we can solve the integral by doing a Laplace transformation from which we get that
\begin{equation}
  \Gamma(z,w)=\frac{W(z)W(w)}{1+g W(z)W(w)} = \frac{W(z)+W(w)}{w+z}\,,
\end{equation}
whose inverse transform gives
\begin{equation}
  \Gamma(t,s)=W(t-s)\,.
\end{equation}
Therefore we get in this case
\begin{equation}
K(t) = - 2\pi g \int_0^{t} dt'\, V(t-t') W(2\pi-t')\,.
\label{Kt}
\end{equation}
Therefore, the ladder contribution reads
\begin{eqnarray}
\label{KBPS}
K(2\pi)
&\!=\!& - 2\pi g \int_0^{2\pi} dt\, \frac{I_0(2\sqrt g(2\pi-t)) I_1(2\sqrt g(2\pi-t))}{\sqrt g (2\pi-t)}
\\
&\!=\!&
-8\pi^2g\,I_0^2(4\pi\sqrt g)
  +2\pi \sqrt g\,I_0(4\pi\sqrt g)I_1(4\pi\sqrt g)
  +8\pi^2g\,I_1^2(4\pi\sqrt g).
  \nonumber
\end{eqnarray}
Ladder resummation gives the exact result in this case.

Analogously, the critical relation that makes \eqref{G(thetatilde)} constant is
\begin{equation}
\cos\gamma = \frac{R_1^2+R_2^2+h^2}{2R_1R_2} \quad  \Rightarrow \quad
\widetilde G(\theta) = g
\end{equation}
Once again, we solve  \eqref{improvedDysonGtilde} doing a Laplace transformation and obtain in this case
\begin{equation}
  \widetilde\Gamma(t,s)=W(t+s)\,.
\end{equation}

With this solution
\begin{equation}
\widetilde K(t) =  2\pi g \int_0^{t} dt'\, V(t-t') W(2\pi+t')\,.
\label{Ktildet}
\end{equation}
Therefore, for the correlator of Wilson loops with the same orientation we get
\begin{eqnarray}
\label{KtildeBPS}
\widetilde K(2\pi)
&\!=\!&
2\pi \sqrt g\,I_0(4\pi\sqrt g)I_1(4\pi\sqrt g).
\end{eqnarray}

The  results \eqref{KBPS} and \eqref{KtildeBPS} were originally obtained from localization, as a two-loop correlator in the Hermitian one-matrix model \cite{Akemann:2001st,Giombi:2009ms}. Details of matrix model results are reviewed in appendix \ref{MatrixModelResults}. The same answer was found in \cite{Correa:2018lyl} by combinatorial methods\footnote{The case of equal orientations was actually not discussed in \cite{Correa:2018lyl}, but the combinatorial counting is identical to the opposite orientation case up to the sign that comes from the constant effective propagator. If the alternating sign in the sum of eq. (67) in \cite{Correa:2018lyl} were removed, the result would have been (\ref{KtildeBPS}).}.

\section{Dyson equation for three loops correlator}

In principle, the same analysis can be extended to account for the connected correlator of any number of concentric circular loops. In order to illustrate how the procedure is generalized, we consider  two representative cases of
three-loop correlators for concentric circles. These connected correlators in the ladder approximation are given by
\begin{align}
 \!\!\!\!\!\left\langle W(\bar C_1)W(C_2) W(C_3)\right\rangle_{\rm conn}\!\!\!\!
 \stackrel{\rm ladd.}{=}&
 \left\langle
 \mathop{\mathrm{tr}}\AU_1(0,2\pi )
 \mathop{\mathrm{tr}}\PU_2(0,2\pi ) \mathop{\mathrm{tr}}\PU_3(0,2\pi )
 \right\rangle_{\rm conn}\!\!\!\!
\\
\!\!\!\!\!\left\langle W(C_1)W(C_2) W(C_3)\right\rangle_{\rm conn}\!\!\!\!
 \stackrel{\rm ladd.}{=}&
 \left\langle
 \mathop{\mathrm{tr}}\PU_1(0,2\pi )
 \mathop{\mathrm{tr}}\PU_2(0,2\pi ) \mathop{\mathrm{tr}}\PU_3(0,2\pi )
 \right\rangle_{\rm conn}\!\!\!\!
\end{align}
As before, the brackets on the right-hand-sides denote Gaussian average with the propagators (\ref{Gaa})--(\ref{G(thetatilde)}).

To compute the first of these quantities, we now define the Green's function
\begin{equation}
M_{123}(t) = N \left\langle
 \mathop{\mathrm{tr}}\AU_1(0,t)
 \mathop{\mathrm{tr}}\PU_2(0,2\pi ) \mathop{\mathrm{tr}}\PU_3(0,2\pi )
 \right\rangle_{\rm conn}
 \label{defM}
\end{equation}
which eventually gives the correlator, when evaluated at $2\pi$
\begin{equation}
\left\langle W(\bar C_1)W(C_2) W(C_3)\right\rangle_{\rm conn}\stackrel{\rm ladd.}{=}
\frac{1}{N} M_{123}(2\pi ).
\end{equation}
The corresponding Dyson equation for $M$ is derived in the appendix~\ref{dyson-appendix} \footnote{We use $\bar a =2,3$ for $a =3,2$ respectively.}:
\begin{align}
 M_{123}(t) = & 2g\int_{0}^{t}dt'\int_{0}^{t'}dt''\,\left[ W(t'-t'')M_{123}(t'') + K_{12}(t'-t'')K_{13}(t'')\right]\nonumber\\
 &+\sum_{a =2}^3 \int_{0}^{t}dt'\int_{0}^{2\pi }d\varphi \,G_{1a}(\varphi -t')
 \Delta_{1a{\bar a}}(t',2\pi|\varphi)
\end{align}
This involves the auxiliary function
\begin{equation}
\Delta_{abc}(t,s|\varphi) = \left\langle \mathop{\mathrm{tr}}[\AU_a(0,t) \PU_b(\varphi,s+\varphi)]\mathop{\mathrm{tr}}\PU_c(0,2\pi )\right\rangle_{\rm conn}
\label{defDelta}
\end{equation}
which itself satisfies another integral equation
\begin{align}
 \!\Delta_{1ab}(t,s|\varphi) & = \widetilde K_{ab}(s)
 \nonumber
 \\
  &+ g\int_{0}^{t}\!dt'\!\int_{0}^{t'}\!dt''
 \left[W(t'-t'')\Delta_{1ab}(t'',s|\varphi) + K_{1b}(t'-t'')\Gamma_{1a}(t'',s|\varphi)\right]
 \nonumber
 \\
 &+\int_{0}^{t}\!dt'\!\int_{0}^{s}\!ds' \,G_{1a}(\varphi+s' -t')
 \left[W(s-s')\Delta_{1ab}(t',s'|\varphi)
 \right.
 \nonumber\\
 & \left.
 \hspace{6cm}+\widetilde K_{ab}(s-s') \Gamma_{1a}(t',s'|\varphi)\right]
 \nonumber
 \\
 &+\int_{0}^{t}\!dt'\!\int_{0}^{2\pi }d\alpha \,G_{1b}(\alpha -t')
 \chi_{1ab}(t',s,2\pi|\varphi,\alpha),
\label{DysonDelta}
\end{align}
in terms of yet another auxiliary function
\begin{equation}
\chi_{1ab}(t,s,u|\varphi,\alpha) = \frac{1}{N}\left\langle \mathop{\mathrm{tr}}\AU_1(0,t)
\PU_a(\varphi ,\varphi +s)\PU_b(\alpha ,\alpha +u)\right\rangle,
\end{equation}
This one finally satisfies an integral equation that closes on itself, provided $W$, $\widetilde\Gamma$ and $\Gamma$ are known:
\begin{align}
\label{Dysonchiimproveds}
 \chi&_{1ab}(t,s,u|\varphi,\alpha) =
 W(t)\widetilde \Gamma_{ab}(s,u|\alpha+u-\varphi)
 \\
 &+\!\int_{0}^{t}\!\!dt'\!\int_{0}^{s}\!\!ds' \,G_{1a}(\varphi+s' -t')
 W(t-t') \Gamma_{1a}(t',s'|\varphi)
 \widetilde \Gamma_{ab}(s-s',u|\alpha+u-\varphi-s')
  \nonumber
 \\
 &+\!\int_{0}^{t}\!\!dt'\!\int_{0}^{u}\!\!du' \,G_{1b}(\alpha +u' -t') W(t-t')W(u-u') \chi_{1ab}(t',s,u'|\varphi,\alpha).\nonumber
\end{align}

\begin{figure}[h]
\center{
\begin{tikzpicture}[scale=0.45,baseline={(0,-0.53)}]
\draw[thick,black!35] (-3.5,2.875) -- (-1,2.4);
\draw[thick,black!35] (-2.5,4.025) -- (-0.4,2.9);
\draw[thick,black!35] (3.5,2.875) -- (1,2.4);
\draw[thick,black!35] (2.5,4.025) -- (0.4,2.9);
\draw[thick,black!35] (-1,0) -- (-.35,2);
\draw[thick,black!35] (1,0) -- (0.35,2);
\draw[black!35] (0,0.4) node {$\cdots$};
\draw[black!35] (2.4,3.25) node {$\ddots$};
\draw[black!35] (-2.4,3.25) node {$\udots$};

\draw[thick,dashed] (-2,4.6) -- (2,4.6);
\draw[thick,dashed] (-2,0) -- (-4,2.3);
\draw[thick,dashed] (2,0) -- (4,2.3);

\fill[red] (-2,0) circle (2pt);
\draw[thick,red,->] (2,0) -- (0,0);
\fill[red] (2,0) circle (2pt);
\draw[thick,red] (-2,0) -- (0.1,0);

\fill[red] (-4,2.3) circle (2pt);
\draw[thick,red] (-3.1,3.335) -- (-2,4.6);
\fill[red] (-2,4.6) circle (2pt);
\draw[thick,red,->] (-4,2.3) -- (-3,3.45);

\fill[red] (4,2.3) circle (2pt);
\draw[thick,red,<-] (3.1,3.335) -- (2,4.6);
\fill[red] (2,4.6) circle (2pt);
\draw[thick,red] (4,2.3) -- (3,3.45);

\fill[yellow] (0,2.3) ellipse (40pt and 25pt);
\draw[black,thick] (0,2.3) ellipse (40pt and 25pt);

\draw[font=\scriptsize] (2,4.9) node {$\alpha$};
\draw[font=\scriptsize] (4.9,2.3) node {$\alpha+u$};
\draw[font=\scriptsize] (-4.4,2.3) node {$\varphi$};
\draw[font=\scriptsize] (-2,4.9) node {$\varphi+s$};
\draw[font=\scriptsize] (-2,-0.4) node {$0$};
\draw[font=\scriptsize] (2,-0.4) node {$t$};

\draw (6.6,2.3) node {$=$};

\end{tikzpicture}
\begin{tikzpicture}[scale=0.45,baseline={(0,-0.53)}]
\draw[thick,black!35] (-3.5,2.875) -- (-1,3);
\draw[thick,black!35] (-2.5,4.025) -- (-0.4,3.5);
\draw[thick,black!35] (3.5,2.875) -- (1,3);
\draw[thick,black!35] (2.5,4.025) -- (0.4,3.5);
\draw[thick,black!35] (-1,0) -- (-.3,1);
\draw[thick,black!35] (1,0) -- (0.3,1);
\draw[black!35] (0,0.25) node {$\cdots$};
\draw[black!35] (2.4,3.4) node {$\ddots$};
\draw[black!35] (-2.4,3.4) node {$\udots$};

\draw[thick,dashed] (-2,4.6) -- (2,4.6);
\draw[thick,dashed] (-2,0) -- (-4,2.3);
\draw[thick,dashed] (2,0) -- (4,2.3);

\fill[red] (-2,0) circle (2pt);
\draw[thick,red,->] (2,0) -- (0,0);
\fill[red] (2,0) circle (2pt);
\draw[thick,red] (-2,0) -- (0.1,0);

\fill[red] (-4,2.3) circle (2pt);
\draw[thick,red] (-3.1,3.335) -- (-2,4.6);
\fill[red] (-2,4.6) circle (2pt);
\draw[thick,red,->] (-4,2.3) -- (-3,3.45);

\fill[red] (4,2.3) circle (2pt);
\draw[thick,red,<-] (3.1,3.335) -- (2,4.6);
\fill[red] (2,4.6) circle (2pt);
\draw[thick,red] (4,2.3) -- (3,3.45);

\fill[magenta2] (0,3.2) ellipse (35pt and 20pt);
\draw[black,thick] (0,3.2) ellipse (35pt and 20pt);

\fill[cyan2] (0,1.2) ellipse (20pt and 14pt);
\draw[black,thick] (0,1.2) ellipse (20pt and 14pt);

\draw[font=\scriptsize] (2,4.9) node {$\alpha$};
\draw[font=\scriptsize] (4.9,2.3) node {$\alpha+u$};
\draw[font=\scriptsize] (-4.4,2.3) node {$\varphi$};
\draw[font=\scriptsize] (-2,4.9) node {$\varphi+s$};
\draw[font=\scriptsize] (-2,-0.4) node {$0$};
\draw[font=\scriptsize] (2,-0.4) node {$t$};

\end{tikzpicture}

\hspace{0.5cm}
\begin{tikzpicture}[scale=0.45,baseline={(0,-0.53)}]
\draw (-5.5,2.3) node {$+$};
\draw[thick,black!35] (-2.7,3.705) -- (-0.2,3.);
\draw[thick,black!35] (-2.1,4.485) -- (0,3.5);
\draw[thick,black!35] (3.5,2.875) -- (1,2.8);
\draw[thick,black!35] (2.5,4.025) -- (0.4,3.4);
\draw[thick,black!35] (-1.5,0) -- (-2.6,1.5);
\draw[thick,black!35] (0.5,0) -- (-1.6,1);

\draw[thick,black!35] (-3.8,2.53) -- (-2,1.);
\draw[thick,black!35] (-3.2,3.22) -- (-1.7,1);

\draw[black!35] (-1,0.15) node {$\cdots$};
\draw[black!35] (2.5,3.3) node {$\ddots$};
\draw[black!35] (-2.2,4) node {$\udots$};
\draw[black!35] (-3.3,2.7) node {$\udots$};

\draw[thick,black!35] (1.2,0) -- (1.45,1);
\draw[thick,black!35] (1.8,0) -- (1.9,1);

\draw[thick,dashed] (-2,4.6) -- (2,4.6);
\draw[thick,dashed] (-2,0) -- (-4,2.3);
\draw[thick,dashed] (2,0) -- (4,2.3);

\draw[thick,dashed,green2] (0.73,0) -- (-3,3.45) ;
\draw[thick,dashed,green2] (1,0) -- (-2.85,3.6215);

\fill[red] (-2,0) circle (2pt);
\draw[thick,red,->] (2,0) -- (1.5,0);

\fill[red] (1,0) circle (2pt);
\draw[thick,red] (1.6,0) -- (1,0);

\fill[red] (0.73,0) circle (2pt);
\draw[thick,red,->] (0.73,0) -- (-0.6,0);
\fill[red] (2,0) circle (2pt);
\draw[thick,red] (-2,0) -- (-0.55,0);

\fill[red] (-4,2.3) circle (2pt);
\fill[red] (-3,3.45) circle (2pt);
\fill[red] (-2.85,3.6215) circle (2pt);
\fill[red] (-2,4.6) circle (2pt);

\draw[thick,red,->] (-4,2.3) -- (-3.5,2.875);
\draw[thick,red] (-3.6,2.76) --(-3,3.45);
\draw[thick,red,->] (-2.85,3.6215) -- (-2.4,4.14);
\draw[thick,red] (-2.45,4.0815)--(-2,4.6);

\fill[red] (4,2.3) circle (2pt);
\draw[thick,red,<-] (3.1,3.335) -- (2,4.6);
\fill[red] (2,4.6) circle (2pt);
\draw[thick,red] (4,2.3) -- (3,3.45);

\fill[magenta2] (-2,1.3) ellipse (20pt and 13pt);
\draw[black,thick] (-2,1.3) ellipse (20pt and 13pt);

\fill[magenta2] (0.5,3.1) ellipse (30pt and 20pt);
\draw[black,thick] (0.5,3.1) ellipse (30pt and 20pt);

\fill[cyan2] (1.7,1.2) ellipse (20pt and 13pt);
\draw[black,thick] (1.7,1.2) ellipse (20pt and 13pt);

\draw[font=\scriptsize] (2,4.9) node {$\alpha$};
\draw[font=\scriptsize] (4.9,2.3) node {$\alpha+u$};
\draw[font=\scriptsize] (-4.4,2.3) node {$\varphi$};
\draw[font=\scriptsize] (-2,4.9) node {$\varphi+s$};
\draw[font=\scriptsize] (-2,-0.4) node {$0$};
\draw[font=\scriptsize] (2,-0.4) node {$t$};
\draw[font=\scriptsize] (0.9,-0.41) node {$t'$};

\draw[font=\scriptsize] (-4,3.45) node {$\varphi+s'$};

\end{tikzpicture}\begin{tikzpicture}[scale=0.45,baseline={(0,-0.53)}]

\draw (-5.5,2.3) node {$+$};

\draw[thick,black!35] (-3.5,2.875) -- (-1,2.4);
\draw[thick,black!35] (-2.5,4.025) -- (-0.4,2.9);
\draw[thick,black!35] (3.2,3.23) -- (0.7,2.5);
\draw[thick,black!35] (2.2,4.37) -- (0.3,3.3);
\draw[thick,black!35] (-1.5,0) -- (-.85,2);
\draw[thick,black!35] (0.5,0) -- (-0.15,2);
\draw[black!35] (-0.5,0.4) node {$\cdots$};
\draw[black!35] (2.2,3.65) node {$\ddots$};
\draw[black!35] (-2.4,3.25) node {$\udots$};

\draw[thick,black!35] (1.2,0) -- (1.8,0.5);
\draw[thick,black!35] (1.8,0) -- (2.05,0.5);

\draw[thick,black!35] (3.65,2.715) -- (3.35,2.2);
\draw[thick,black!35] (3.9,2.415) -- (3.5,2.06);

\draw[thick,dashed] (-2,4.6) -- (2,4.6);
\draw[thick,dashed] (-2,0) -- (-4,2.3);
\draw[thick,dashed] (2,0) -- (4,2.3);

\draw[thick,dashed,green2] (0.73,0) -- (3.4,3);
\draw[thick,dashed,green2] (1,0) -- (3.55,2.83);

\fill[red] (-2,0) circle (2pt);
\draw[thick,red,->] (2,0) -- (1.5,0);

\fill[red] (1,0) circle (2pt);
\draw[thick,red] (1.6,0) -- (1,0);

\fill[red] (0.73,0) circle (2pt);
\draw[thick,red,->] (0.73,0) -- (-0.6,0);
\fill[red] (2,0) circle (2pt);
\draw[thick,red] (-2,0) -- (-0.55,0);

\fill[red] (-4,2.3) circle (2pt);
\draw[thick,red] (-3.1,3.335) -- (-2,4.6);
\fill[red] (-2,4.6) circle (2pt);
\draw[thick,red,->] (-4,2.3) -- (-3,3.45);

\fill[red] (2,4.6) circle (2pt);
\fill[red] (3.4,3) circle (2pt);
\fill[red] (3.55,2.83) circle (2pt);
\fill[red] (4,2.3) circle (2pt);

\draw[thick,red,->] (2,4.6) -- (2.9,3.565);
\draw[thick,red] (2.8,3.68) -- (3.4,3);
\draw[thick,red,->] (3.55,2.83) -- (3.85,2.485);
\draw[thick,red]  (3.75,2.6) -- (4,2.3);

\fill[yellow] (-0.5,2.8) ellipse (40pt and 25pt);
\draw[black,thick] (-0.5,2.8) ellipse (40pt and 25pt);

\fill[cyan2] (1.95,0.55) ellipse (8pt and 6pt);
\draw[black,thick] (1.95,0.55) ellipse (8pt and 6pt);

\fill[cyan2] (3.25,2) ellipse (8pt and 6pt);
\draw[black,thick] (3.25,2) ellipse (8pt and 6pt);

\draw[font=\scriptsize] (2,4.9) node {$\alpha$};
\draw[font=\scriptsize] (4.9,2.3) node {$\alpha+u$};
\draw[font=\scriptsize] (-4.4,2.3) node {$\varphi$};
\draw[font=\scriptsize] (-2,4.9) node {$\varphi+s$};
\draw[font=\scriptsize] (-2,-0.4) node {$0$};
\draw[font=\scriptsize] (2,-0.4) node {$t$};

\draw[font=\scriptsize] (0.9,-0.41) node {$t'$};
\draw[font=\scriptsize] (4.45,3.03) node {$\alpha+u'$};

\end{tikzpicture}
\caption{Diagrammatic interpretation of the integral equation (\ref{Dysonchiimproveds})}
\label{figchi}
}
\end{figure}
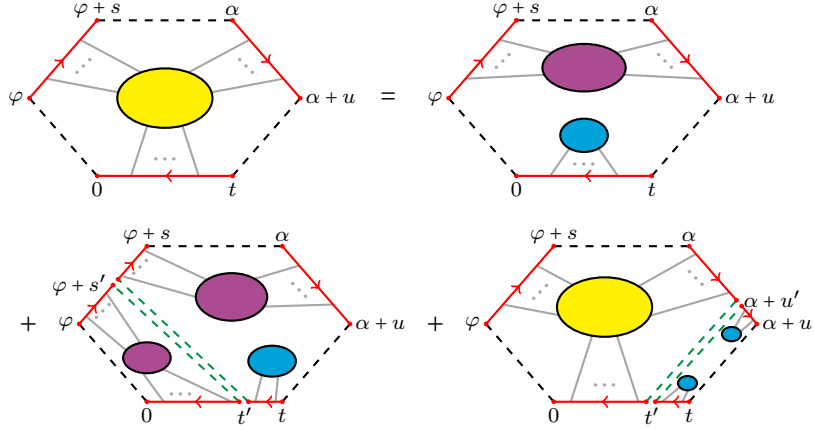

We can interpret diagrammatically this equation through figure \ref{figchi}. The first term comes from diagrams with no connecting propagator from the loop 1. In the remaining terms, $t'$ stands for the rightmost point in $\AU_1(0,t)$ with a connecting propagator. Thus, from the propagators in between $t'$ and $t$ we have a $W(t-t')$ factor. Between $0$ and $t'$ we do have connecting propagators and in the planar approximation
we get the second and third terms when $t'$ connects with a point in $\PU_a(\varphi,\varphi+s)$ and a point in $\PU_b(\alpha,\alpha+u)$ respectively.

When the three loops have the same orientation we define
\begin{align}
	\widetilde M_{abc}(t) & = N \langle \mathop{\mathrm{tr}}\PU_a(0,t)\mathop{\mathrm{tr}}\PU_b(0,2\pi )
\mathop{\mathrm{tr}}\PU_c(0,2\pi )\rangle_{\rm conn},
	\label{Mtilde}
\\
\widetilde \Delta_{abc} (t,s|\varphi ) &= \langle\mathop{\mathrm{tr}}\PU_a(0,t)\PU_b(\varphi-s,\varphi)\mathop{\mathrm{tr}}\PU_c(0,2\pi )\rangle_{\rm conn},\label{Deltatilde}
\\
\label{Gammatilde}
	\widetilde \chi_{abc} (t,s,u|\varphi,\alpha ) &= \frac{1}{N}\langle\mathop{\mathrm{tr}}\PU_a(0,t)\PU_b(\varphi-s,\varphi)\PU_c(\alpha-u,\alpha)\rangle.
\end{align}
for which we obtain the following set of integral equations
\begin{align}
 \widetilde M_{123}(t) = & 2g\int_{0}^{t}dt'\int_{0}^{t'}dt''\,\left[ W(t'-t'')\widetilde M_{123}(t'') + \widetilde K_{12}(t'-t'') \widetilde K_{13}(t'')\right]\nonumber\\
 &+\sum_{a =2}^3 \int_{0}^{t}dt'\int_{0}^{2\pi }d\varphi \,\widetilde G_{1a}(\varphi -t')
 \widetilde\Delta_{1a{\bar a}}(t',2\pi|\varphi)
\end{align}
\begin{align}
 \!\widetilde\Delta_{1ab}(t,s|\varphi) & = \widetilde K_{ab}(s)
 \nonumber
 \\
  &+ g\int_{0}^{t}\!dt'\!\int_{0}^{t'}\!dt''
 \left[W(t'-t'')\widetilde\Delta_{1ab}(t'',s|\varphi) + \widetilde K_{1b}(t'-t'')\widetilde \Gamma_{1a}(t'',s|\varphi)\right]
 \nonumber
 \\
 &+\int_{0}^{t}\!dt'\!\int_{0}^{s}\!ds' \,\widetilde G_{1a}(\varphi-s' -t')
 \left[W(s-s')\widetilde\Delta_{1ab}(t',s'|\varphi)
 \right.
 \nonumber\\
 & \left.
 \hspace{6cm}+\widetilde K_{ab}(s-s') \widetilde\Gamma_{1a}(t',s'|\varphi)\right]
 \nonumber
 \\
 &+\int_{0}^{t}\!dt'\!\int_{0}^{2\pi }d\alpha \,\widetilde G_{1b}(\alpha -t')
 \widetilde\chi_{1ab}(t',s,2\pi|\varphi,\alpha),
 \label{DysontildeDelta}
\end{align}
\begin{align}
\nonumber
 \widetilde\chi&_{1ab}(t,s,u|\varphi,\alpha) =
 W(t) \widetilde\Gamma_{ab}(s,u|\varphi-\alpha-s)
 \\
 &+\!\int_{0}^{t}\!\!dt'\!\int_{0}^{u}\!\!du' \,\widetilde G_{1b}(\alpha-u' -t')
 W(t-t') \widetilde\Gamma_{1b}(t',u'|\alpha)
 \widetilde \Gamma_{ab}(s,u-u'|\varphi-\alpha+u'-s)
  \nonumber
 \\
 &+\!\int_{0}^{t}\!\!dt'\!\int_{0}^{s}\!\!ds' \,\widetilde G_{1a}(\varphi -s' -t') W(t-t')W(s-s') \widetilde\chi_{1ab}(t',s',u|\varphi,\alpha).\label{Dysontildechiimproved}
\end{align}

These equations completely determine the  ladder contribution to the three-loop correlator. In the next section we show how to solve them for the BPS configurations, when the parameters are adjusted to make all propagators constant.

\subsection{Solution for the BPS configurations}

In the BPS case $G_{1a}=-g$ and the dependence of $\varphi$ and $\alpha$ drops from (\ref{Dysonchiimproveds}). The Laplace transformation of this integral equation gives
\begin{equation}
\chi(z,v,w) = \frac{\widetilde\Gamma(v,w)}{z}[1-g\Gamma(z,v)]
+\frac{g}z[W(z)-W(w)]\chi(z,v,w).
\end{equation}
Thus,
\begin{align}
\chi(z,v,w)
&= \frac{\widetilde\Gamma(v,w)[1-g\Gamma(z,v)]}{z-g[W(z)-W(w)]}= \frac{W(w)-W(v)}{(v-w)(z+w)}+\frac{W(z)+W(v)}{(z+w)(z+v)},
\end{align}
which is the triple Laplace transform of $W$ with argument $t-s-u$. Thus, we simply have
\begin{equation}
\chi(t,s,u) = W(t-s-u).
\end{equation}

Let us now turn to the auxiliary function  $\Delta_{1ab}(t,s|\varphi)$.  Since in this case $\Delta_{123}$ and $\Delta_{132}$ are equal and independent of $\varphi$, we will denote them as $\Delta(t,s)$. The Laplace transform of its integral equation gives
\begin{align}
\Delta(z,v) =& \frac{\widetilde{K}(v)}{z}+\frac{g}{z}[W(z)-W(v)]\Delta(z,v)
+\frac{g}{z}[K(z)-\widetilde K(v)]\Gamma(z,v)
\nonumber\\
&-\frac{2\pi g}{z} \frac{J(z)+\widetilde{J}(v)}{z+v},
\end{align}
where $J(z)$  and $\widetilde J(z)$ are the Laplace transforms of $J(t) = W(t-2\pi)$ and  $\widetilde J(t) = W(t+2\pi)$ respectively.
If we further use that
\begin{equation}
K(z) = -\frac{2\pi g J(z)}{z-2g W(z)},\qquad   \widetilde K(z) = +\frac{2\pi g \widetilde J(z)}{z-2g W(z)},
\end{equation}
we obtain
\begin{align}
\Delta(z,v)
=& \frac{K(z)+\widetilde K (v)}{z+v},
\end{align}
which means that
\begin{equation}
\Delta(t,s) = K(t-s).
\end{equation}

Finally, with this result we get for the BPS connected correlator of three loops
\begin{equation}
M(2\pi) = 2g \int_0^{2\pi}\! dt\! \int_0^{t}\! dt' V(2\pi-t)K(t-t')K(t')
-4\pi g \int_0^{2\pi}\! dt V(2\pi-t)K(t-2\pi),
\end{equation}
which using the exact result (\ref{Kt}) gives
\begin{equation}\label{eme}
M(2\pi) = (2\pi\sqrt{g})^3\left(I_1^3(4\pi\sqrt g)
  -I_0(4\pi\sqrt g)^2I_1(4\pi\sqrt g)\right),
\end{equation}
in agreement with the direct calculation in the  matrix model (\ref{CbarCC}).

In the critical case of $\widetilde G_{ab} = g$, the integral equations for the case of three loops with the same orientation can be solved by doing Laplace transformations. We obtain in this case
\begin{align}
\widetilde\chi(t,s,u) & = W(t+s+u),
\\
\widetilde\Delta(t,s) & = \widetilde K(t+s),
\end{align}
and from these
\begin{align}\label{emet}
\widetilde M(2\pi) = & 2g \int_0^{2\pi}\! dt\! \int_0^{t}\! dt' V(2\pi-t)\widetilde K(t-t')\widetilde K(t')
+4\pi g \int_0^{2\pi}\! dt V(2\pi-t) \widetilde K(2\pi+t)\nonumber
\\
= & (2\pi\sqrt{g})^3\left(I_1^3(4\pi\sqrt g)
  +3I_0(4\pi\sqrt g)^2I_1(4\pi\sqrt g)\right),
\end{align}
in agreement with the matrix model result (\ref{CCC}).

The results given in eqs.\ \eqref{KBPS}, \eqref{KtildeBPS}, \eqref{eme} and \eqref{emet} for BPS configurations can be related to the results of connected correlators of more general Wilson loops also computable in terms of matrix models. More precisely, using multi-matrix models  \cite{Giombi:2009ms,Giombi:2012ep} it is possible to obtain the connected correlators of the $\frac{1}{8}$ BPS Wilson loops supported in arbitrary curves on a $S^2$ \cite{Drukker:2007dw,Drukker:2007yx,Drukker:2007qr}. In particular  eq.\ (8.79) of \cite{Giombi:2009ms} reproduces our eqs.\ \eqref{KBPS} and \eqref{KtildeBPS}, whereas eq.\ (4.39) of \cite{Giombi:2012ep} reproduces our eqs.\ \eqref{eme} and \eqref{emet}  when the $\frac{1}{8}$ BPS Wilson loops are taken to be coincident.

\section{Conclusions}

We have studied correlators of circular Wilson loops in the ladder approximation.
For the supersymmetric configurations, no approximation is made by restricting to ladders and their resummation yields exact results for Wilson loop correlators.
Moreover, resummation of ladders in this case is a combinatorial problem accounted for by the Gaussian matrix model.
More generally, ladder resummation cannot be rigorously justified, but still results in a qualitative agreement with expectations from string theory. In particular, the  phase diagram of the string-breaking transition is qualitatively similar to the one obtained from minimal area law in $AdS_{5}\times S^{5}$. The numerical details differ because ladders do not account for all possible contributions at large 't~Hooft coupling.

Recently found connections between Dyson equations for ladder diagrams and the AdS/CFT integrability \cite{Gromov:2016rrp,Kim:2017sju,Cavaglia:2018lxi} is suggestive of a deeper mathematical structure behind ladder resummation. It would be extremely interesting to understand how  integrable structures arise in Wilson loop correlators studied in this paper. The first steps in this direction have been made in \cite{Giombi:2018qox}.

\subsection*{Acknowledgements}
The work of K.~Z. was supported by the ERC advanced grant No 341222, by the Swedish Research Council (VR) grant
2013-4329, by the grant ``Exact Results in Gauge and String Theories" from the Knut and Alice Wallenberg foundation, and by RFBR grant 18-01-00460 A. K.Z. was partially supported by the Simons Foundation
under the program Targeted Grants to Institutes (the Hamilton Mathematics
Institute). The work of A.~R.~F. was partially provided by the Spanish MINECO under
projects ``Holography and QCD in extreme conditions" and MDM-2014-0369 of ICCUB (Unidad de Excelencia `Mar\'\i a de Maeztu'). The work of D.~C. was supported in part by grants PIP 0681, and PID B\'usqueda de
nueva F\'\i sica and UNLP X850.

\appendix

\section{Average number of propagators}
\label{average-l}

Consider the perturbative expansion of a Wilson loop expectation value (or a correlator, at this level the difference is immaterial):
\begin{equation}
\left\langle  W(C) \right\rangle=\sum_{\ell}^{}w_{\ell}\lambda ^{\ell}.
\end{equation}
The order of perturbation theory $\ell$ counts the number of loops, which for ladders coincides with the number of propagators. The  dominant contribution comes from diagrams of order
\begin{equation}
 \bar{\ell}=\frac{\sum\limits_{\ell}^{}\ell w_{\ell}\lambda ^{\ell}}{\sum\limits_{\ell}^{}w_{\ell}\lambda ^{\ell}}=\lambda \,\frac{\partial }{\partial \lambda }\,\ln
 \left\langle  W(C) \right\rangle.
\end{equation}

At strong coupling, the AdS/CFT correspondence predicts an exponential
growth of the correlator:
\begin{equation}
 \ln\left\langle  W(C) \right\rangle\simeq  \frac{A_{{\rm r}}}{2\pi }\,\sqrt{\lambda },
\end{equation}
where $A_{{\rm r}}$ is minus the regularized area in $AdS_{5}\times S^{5}$ (one can show that $A_{{\rm r}}>0$). The order at which diagrams contribute most thus  grows as the square root of the coupling:
\begin{equation}
 \bar{\ell}\simeq  \frac{A_{{\rm r}}}{4\pi }\,\sqrt{\lambda }\,.
\end{equation}
The ladder approximation shares the square-root exponential scaling with the exact answer \cite{Erickson:1999qv,Erickson:2000af,Zarembo:2001jp,Klebanov:2006jj}. The diagram counting therefore is the same up to a numeric coefficient.

\section{Derivation of Dyson equations}\label{dyson-appendix}
The ordered exponentials (\ref{PUAU}), used to define the Green's functions, are solutions to the following recursion relations:
\begin{eqnarray}
 \PU_a(t_1,t_2)&=&\mathbbm{1}+\int_{t_1}^{t_2}dt\,\PU_a(t_1,t){\cal O}_a(t),
\\
\label{AU-rec}
  \AU_a(t_1,t_2)&=&\mathbbm{1}+\int_{t_1}^{t_2}dt\,{\cal O}_a(t)\AU_a(t_1,t).
\end{eqnarray}

The Dyson equations follow from these recursion relations upon applying Wick's theorem:
\begin{equation}
 \left\langle {\cal O}_a\mathcal{F}({\cal O} )\right\rangle=
 \widehat{{\cal O}_a{\cal O}_b}\,
 \left\langle \frac{\partial \mathcal{F}}{\partial {\cal O}_b}\right\rangle,
\end{equation}
with subsequent use of the large-$N$ factorization. Wick's theorem applies because fields $\mathcal{O}_a(t)$ are Gaussian.

For example, starting with a single trace of an ordered exponential, we have
\begin{eqnarray}
 &&\left\langle\mathop{\mathrm{tr}}\AU_a(0,t)\right\rangle
 =N+\int_{0}^{t}dt'\,\left\langle \mathop{\mathrm{tr}}{\cal O}_a(t')\AU_1(0,t')\right\rangle
\nonumber \\&&=
 N+\frac{g}{N}\int_{0}^{t}dt'\,\int_{0}^{t'}dt''\,\left\langle \mathop{\mathrm{tr}}\AU_a(0,t'')\mathop{\mathrm{tr}}\AU_a(t'',t')\right\rangle,
\end{eqnarray}
where (\ref{AU-rec}) is used in the first equality and Wick's theorem in the second one. Finally, applying large-$N$ factorization and recalling that
\begin{equation}\label{defW}
 W(t) =\frac{1}{N} \left\langle\mathop{\mathrm{tr}}\AU_a(0,t)\right\rangle,
\end{equation}
we get an integral equation for $W(t)$:
\begin{equation}\label{DysonW}
 W(t)=1+g\int_{0}^{t}dt'\int_{0}^{t'}dt''\,W(t'-t'')W(t'').
\end{equation}
This is the loop equation for the Gaussian one-matrix model \cite{Migdal:1984gj},\cite{Makeenko:1991tb}, and can be easily solved by a Laplace transform:
\begin{equation}
W(t) = \frac{1}{\sqrt g t} I_1\left(2\sqrt g t\right),
\end{equation}
where $I_1$ is the modified Bessel function.

Applying the same chain of arguments, we can derive the integral equations that describe the ladder contribution to the connected two-loop correlator. Using the relation (\ref{AU-rec}) and Wick's theorem on a correlator of two ordered exponentials we get
\begin{eqnarray}
 &&\left\langle \mathop{\mathrm{tr}}\AU_a(0,t)\mathop{\mathrm{tr}}\PU_b(0,2\pi )\right\rangle
 = N \left\langle \mathop{\mathrm{tr}}\PU_b(0,2\pi )\right\rangle
\nonumber \\ &&
 +\frac{g}{N}\int_{0}^{t}dt'\int_{0}^{t'}dt''\,
 \left\langle \mathop{\mathrm{tr}}\AU_a(0,t'')\mathop{\mathrm{tr}}\AU_a(t'',t')\mathop{\mathrm{tr}}\PU_b(0,2\pi )\right\rangle
\nonumber \\ &&
 +\frac{1}{N}\int_{0}^{t}dt'\int_{0}^{2\pi }d\varphi \,G(\varphi -t')
 \left\langle \mathop{\mathrm{tr}}\AU_a(0,t')\PU_b(\varphi ,\varphi +2\pi )\right\rangle.
\end{eqnarray}
Applying large-$N$ factorization in the second line and taking the connected part of the correlator  we get an equation that can be expressed in terms of the Green's functions  $K$ and $\Gamma $ defined in (\ref{K})-(\ref{Gamma}):
\begin{equation}
 K_{ab}(t)=2g\int_{0}^{t}dt'\int_{0}^{t'}dt''\,W(t'-t'')K_{ab}(t'')
 +\int_{0}^{t}dt'\int_{0}^{2\pi }d\varphi \,G(\varphi -t')\Gamma_{ab} (t',2\pi |\varphi ).
\end{equation}

The auxiliary function $\Gamma_{ab}(t,s|\varphi )$ satisfies a closed Dyson equation \cite{Zarembo:2001jp}. Here we rederive it applying the relation (\ref{AU-rec}) and Wick's theorem to the defining expectation value of  $\Gamma$ (\ref{Gamma}):
\begin{eqnarray}
 &&\left\langle \mathop{\mathrm{tr}}\AU_a(0,t)\PU_b(\varphi ,\varphi +s)\right\rangle=\left\langle \mathop{\mathrm{tr}}\PU_b(\varphi ,\varphi +s)\right\rangle
\nonumber \\ &&
 +\frac{g}{N}\int_{0}^{t}dt'\int_{0}^{t'}dt''\,\left\langle \mathop{\mathrm{tr}}\AU_a(0,t'')\PU_b(\varphi ,\varphi +s)\mathop{\mathrm{tr}}\AU_a(t'',t')\right\rangle
\nonumber \\ &&
 +\frac{1}{N}\int_{0}^{t}dt'\int_{0}^{s}ds'\,G(\varphi +s'-t')
\nonumber \\&&\times
 \left\langle \mathop{\mathrm{tr}}\AU_a(0,t')\PU_b(\varphi ,\varphi +s')\mathop{\mathrm{tr}}\PU_b(\varphi +s',\varphi +s)\right\rangle.
\end{eqnarray}
The double-trace correlators factorize in the large-$N$ limit, and we get a closed equation for $\Gamma_{ab}$:
\begin{eqnarray}\label{asymDysonGamma}
 \Gamma_{ab}(t,s|\varphi )&=&W(s)+g\int_{0}^{t}dt'\int_{0}^{t'}dt''\,W(t'-t'')\Gamma_{ab}(t'',s|\varphi )
\nonumber \\
&&+\int_{0}^{t}dt'\int_{0}^{s}ds'\,G(\varphi +s'-t')W(s-s')\Gamma_{ab} (t',s'|\varphi ).
\end{eqnarray}

This equation can be brought to a more symmetric form with the help of the following argument. Consider an integral equation
\begin{equation}
f(t) = g\int_{0}^{t} dt'   \int_{0}^{t'}dt'' \,W(t'-t'')f (t'')+
\int_{0}^{t} dt'\, j(t'),
\label{identity1}
\end{equation}
where $f(t)$ is an unknown and $j(t)$ is given.
Due to the fact that $W(t)$ satisfies (\ref{DysonW}), equation (\ref{identity1}) is solved by
\begin{equation}
 f(t)=\int_{0}^{t}dt'\,W(t-t')j(t'),
  \label{identity2}
\end{equation}
as can be checked by direct substitution. Applying this result to the equation (\ref{asymDysonGamma}) brings the latter to a symmetric form  quoted in the main text as (\ref{improvedDysonG}).

For the connected three-loop correlator we need to derive an integral equation for the triple-trace correlator. Using (\ref{AU-rec}) and Wick's theorem we get \footnote{We use $\bar a =2,3$ for $a =3,2$ respectively.}
\begin{align}
 &\left\langle \mathop{\mathrm{tr}}\AU_1(0,t)\mathop{\mathrm{tr}}\PU_2(0,2\pi ) \mathop{\mathrm{tr}}\PU_3(0,2\pi )\right\rangle
 = N \left\langle \mathop{\mathrm{tr}}\PU_2(0,2\pi ) \mathop{\mathrm{tr}}\PU_3(0,2\pi )\right\rangle
 \\ &
 +\frac{g}{N}\int_{0}^{t}dt'\int_{0}^{t'}dt''\,
 \left\langle \mathop{\mathrm{tr}}\AU_1(0,t'')\mathop{\mathrm{tr}}\AU_1(t'',t')\mathop{\mathrm{tr}}\PU_2(0,2\pi )\mathop{\mathrm{tr}}\PU_3(0,2\pi )\right\rangle
\nonumber \\ &
 +\frac{1}{N}\sum_{a =2}^3 \int_{0}^{t}dt'\int_{0}^{2\pi }d\varphi \,G_{1a}(\varphi -t')
 \left\langle \mathop{\mathrm{tr}}[\AU_1(0,t')\PU_a(\varphi ,\varphi +2\pi )]\mathop{\mathrm{tr}}\PU_{\bar a}(0,2\pi )\right\rangle.\nonumber
\end{align}

Applying large-N factorization and keeping the connected part, we get an equation for the correlator that can be expressed in terms of $M$ and $\Delta$ defined in (\ref{defM}) and (\ref{defDelta})
\begin{align}
 M_{123}(t) = & 2g\int_{0}^{t}dt'\int_{0}^{t'}dt''\,\left[ W(t'-t'')M_{123}(t'') + K_{12}(t'-t'')K_{13}(t'')\right]\nonumber\\
 &+\sum_{a =2}^3 \int_{0}^{t}dt'\int_{0}^{2\pi }d\varphi \,G_{1a}(\varphi -t')
 \Delta_{1a{\bar a}}(t',2\pi|\varphi)
\end{align}

\section{Matrix models for BPS correlators}
\label{MatrixModelResults}
The correlators of BPS circular Wilson loops can be obtained from a Gaussian matrix model with partition function
\begin{equation}
{\cal Z} = \int dM e^{-\frac{N}{2g}{\rm tr}(M^2)}.
\end{equation}

The connected correlators are computed from,
\begin{equation}
W(t_1,\cdots,t_k) = N^{k-2}\langle {\rm tr}e^{t_1 M} \cdots {\rm tr}e^{t_k M}\rangle_{\rm conn},
\end{equation}
whose Laplace transforms are the $k$-point resolvents
\begin{equation}
W(z_1,\cdots,z_k) = N^{k-2}\left\langle {\rm tr}\frac{1}{z_1-M} \cdots {\rm tr}\frac{1}{z_k-M}\right\rangle_{\rm conn},
\end{equation}

In \cite{Akemann:2001st} the first $k$-point resolvents are explicitly presented. To leading order in the large $N$ limit
\begin{align}
  W(z_1) &= \frac{1}{2g}(z_1-\sqrt{z_1^2-4g})
  \\
  W(z_1,z_2)&=\frac{1}{2(z_1-z_2)}
  \left(\frac{z_1 z_2-4g}{\sqrt{(z_1^2-4g)(z_2^2-4g)}}-1\right)
  \\
  W(z_1,z_2,z_3)&= \frac{2g^2\left(z_1 z_2+z_1 z_3+z_2 z_3+4g\right)}{\left[(z_1^2-4g)(z_2^2-4g)(z_3^2-4g)\right]^{\frac32}}.
\end{align}
Upon inverse Laplace transformation we obtain
\begin{align}
W(t_1)&=\frac{1}{\sqrt{g} t_1} I_1(2\sqrt{g} t_1)
\label{w1}
\\
  W(t_1,t_2)&=\sqrt{g}\frac{t_1 t_2}{t_1+t_2}\left[I_0(2\sqrt{g}t_1)I_1(2\sqrt{g}t_2)+I_1(2\sqrt{g}t_1)I_0(2\sqrt{g}t_2)\right]
  \label{w2}
  \\
   W(t_1,t_2,t_3)&=g^{\frac32}\,t_1 t_2 t_3
  \left[I_1(2\sqrt{g}t_1)I_0(2\sqrt{g}t_2)I_0(2\sqrt{g}t_3)\right.
    \nonumber
  \\
  &\hspace{2cm}+I_0(2\sqrt{g}t_1)I_1(2\sqrt{g}t_2)I_0(2\sqrt{g}t_3)
  \nonumber
  \\
  & \hspace{2cm}+I_0(2\sqrt{g}t_1)I_0(2\sqrt{g}t_2)I_1(2\sqrt{g}t_3)
  \nonumber\\
  &\hspace{2cm}+\left.I_1(2\sqrt{g}t_1)I_1(2\sqrt{g}t_2)I_1(2\sqrt{g}t_3)\right]
  \label{w3}
\end{align}

From (\ref{w2}) we obtain the connected two-loop correlators: $W(-2\pi,2\pi)$ gives the correlator in the case of loops with opposite orientation while $W(2\pi,2\pi)$ gives the correlator for loops with the same orientation
\begin{align}
\left\langle W(\bar C_1)W(C_2)\right\rangle_{\rm BPS} = & -8\pi^2g\,I_0^2(4\pi\sqrt g)
  +2\pi \sqrt g\,I_0(4\pi\sqrt g)I_1(4\pi\sqrt g)\nonumber\\
  &+8\pi^2g\,I_1^2(4\pi\sqrt g)
  \label{CbarC}
  \\
\left\langle W(C_1)W(C_2)\right\rangle_{\rm BPS} = & 2\pi \sqrt g\,I_0(4\pi\sqrt g)I_1(4\pi\sqrt g)
\label{CC}
\end{align}

Similarly, from (\ref{w3}) we obtain the connected three-loop correlators. $W(-2\pi,2\pi,2\pi)$ gives the correlator in the case in which one of the loops has opposite orientation, while $W(2\pi,2\pi,2\pi)$ gives the correlator for the three loops with the same orientation.

\begin{align}
\left\langle W(\bar C_1)W(C_2)W(C_3)\right\rangle_{\rm BPS} = & (2\pi\sqrt{g})^3\left(I_1^3(4\pi\sqrt g)
  -I_0(4\pi\sqrt g)^2I_1(4\pi\sqrt g)\right)
  \label{CbarCC}
  \\
\left\langle W(C_1)W(C_2)W(C_3)\right\rangle_{\rm BPS} = &
(2\pi\sqrt{g})^3\left(I_1^3(4\pi\sqrt g)
  +3I_0(4\pi\sqrt g)^2I_1(4\pi\sqrt g)\right)
\label{CCC}
\end{align}

\bibliographystyle{nb}

\end{document}